\input{psfig}
\documentclass[referee]{aa}
\thesaurus{12.03.4; 12.12.1; 11.06.1}
\title{On the evolution of aspherical perturbations in the universe:  
an analytical model}
\author{A. Del Popolo\inst{1,2,3}}
\offprints{A. Del Popolo - {\it E-mail adelpop@unibg.it}}
\institute{$^1$ Dipartimento di Matematica, Universit\`{a} Statale di Bergamo,
via dei Caniana, 2 - I 24129 Bergamo, ITALY \\
$^2$ Feza G\"ursey Institute, P.O. Box 6 \c Cengelk\"oy, Istanbul, Turkey \\
$^3$ Bo$\breve{g}azi$\c{c}i University, Physics Department,
80815 Bebek, Istanbul, Turkey}
\titlerunning{On the evolution of aspherical perturbations}
\authorrunning{A. Del Popolo}
\date{}
\begin{document}
\maketitle
\begin{abstract}
I study the role of shear fields
on the evolution of density perturbations
by using an analytical approximate solution for the equations of
motion of homogeneous ellipsoids embedded in a homogeneous
background.
The equations of motion of a homogeneous ellipsoid (Icke 1973;
White \& Silk 1979 (hereafter WS)) are modified in order to take
account of the tidal field as done in Watanabe 1993 and then are
integrated analytically, similarly to what done in WS. The
comparison of the analytical solution with numerical simulations
shows that it is a good approximation to the numerical one. This
solution is used to study the evolution of the configuration of
the ellipsoids, to calculate the evolution of the density contrast
and that of the axial peculiar velocity of the ellipsoids for
several values of the amplitude of the external tidal field,
and compared again with numerical simulations. In order to
calculate the evolution of the density contrast at turn-around and
collapse velocity at the epoch of collapse, as a function of the
ratio of the initial value of the semi-axes, I use the previously
obtained approximate solution to modify the analytical model
proposed by Barrow \& Silk (1981) for the ellipsoids evolution in
the non-linear regime. The density contrast at turn-around and the
collapse velocity are found to be reduced with respect to that
found by means of the spherical model. The reduction increases
with increasing strength of the external tidal field and with
increasing initial asymmetry of the ellipsoids. These last
calculations are also compared with numerical solutions and they
are again in good agreement with the numerical ones.
\end{abstract}
\keywords{cosmology: theory - large scale structure of universe - galaxies:
formation}
\section{Introduction}
Most clusters and superclusters, similarly to elliptical galaxies,
are not spherical and their shape is not due to rotation (Rood et
al. 1972; Gregory \& Tifft 1976; Dressler 1981). This deviation
from spherical symmetry is seen wherever large-scale structures
can be unambiguously delineated. Going towards very large scales
this tendency to the 'breaking of symmetry' takes the form of
strong inhomogeneity and subclustering. For example, the Local
Supercluster (LSC) has an appreciable flatness of the axial ratio,
1:6-1:9 (see Tully 1982).
On smaller scales, clusters of galaxies are often highly elongated
and galaxies are significantly aspherical. In the framework of the
gravitational instability, in which structures formed by the
gravitational amplification of small primeval inhomogeneities,
this lack of spherical symmetry should be somehow expected. In
fact, the perturbations that gave rise to the formation of
structures are alike to have been initially aspherical (Barrow \&
Silk 1981 (hereafter BS); Peacock \& Heavens 1985; Bardeen et al.
1986) and asphericities are then amplified during gravitational
collapse (Lin et al. 1965; Icke 1973; BS). The elongations are
probably due to a velocity anisotropy of the galaxies (Aarseth \&
Binney 1978). According to Binney \& Silk (1979) and to
Salvador-Sol\'e \& Solanes (1993) the elongation of clusters
originates in the tidal distortion by neighboring protoclusters.
In particular, Salvador-Sol\'e \& Solanes (1993) found that the
main distortion on a cluster is produced by the nearest
neighboring ones having more than 45 galaxies and the same
model can explain the observed alignment between neighboring
clusters (Binggeli 1982; Oort 1983; Rhee \& Katgert 1987; Plionis
1994) and that between clusters and their first ranked galaxy
(Carter \& Metcalfe 1980; Dressler 1981; Binggeli 1982; Rhee \&
Katgert 1987; Tucker \& Peterson 1988; van Kampen \& Rhee 1990;
Lambas et al. 1990; West 1994). \footnote{Clusters elongations and
alignment could be also explained by means of Zeldovich's (1978)
``pancakes" theory of cluster formation but this top-down formation
model is probably ruled out for several well known reasons
(Peebles 1993).} As previously reported, another characteristic
found by some of the studies previously quoted (e.g., Binggeli
1982), is a tendency of clusters through alignment, seen for the
first time in two-dimensional distribution of Abell Clusters
(Binggeli 1982). The evidences reported suggest that the
anisotropy of the external density field plays an important role
in the evolution of clusters or superclusters. To be more precise,
we should tell that
the role of shear on structure formation is of fundamental
importance. Shear on a density perturbation can be produced by the
intrinsic asphericity of the perturbation itself (internal shear)
or it can be due to the interaction of the perturbation with the
neighboring ones (external shear) \footnote{
the term ``internal shear" has the meaning of ``intrinsic asphericity"
while 
``external shear" that of ``non-spherical
external force".
}
The role of external shear on structure formation was studied theoretically
in Binney \& Silk (1979) and Palmer (1983). In particular, Palmer (1983),
studied the gravitational collapse of homogeneous prolate spheroids under
the influence of a single attractor.
The importance of the external shear
for structure formation was recognized by Hoffman (1986a, 1989).
In particular Hoffman (1986a,1989), using the quasi-linear (QL)
approximation (Zel'dovich 1970; Zel'dovich \& Novikov 1983),
showed that the shear affects the dynamics of collapsing objects
and leads to infall velocities that are larger than in the case
of non-shearing ones.
As a consequence, according to Hoffmann (1986a, 1989), a general
mass element collapse faster than a spherical one. Evrard \& Crone
(1992), Bertschinger \& Jain (1994) and Monaco (1995), arrived at
similar conclusions. Bertschinger \& Jain (1994) put this result
in theorem form according to which spherical perturbations are
the slowest in collapsing.

On the actuality of the Bertschinger \& Jain's collapse theorem
there is not fully agreement in literature. For example, according
to the previrialization conjecture (Peebles \& Groth 1976, Davis
\& Peebles 1977, Peebles 1990), initial asphericities and tidal
interactions between neighboring density fluctuations induce
significant non-radial motions which oppose the collapse. This
kind of conclusion was supported by BS, Szalay \& Silk (1983),
Villumsen \& Davis (1986), Bond \& Myers (1993a,b)
and Lokas et al. (1996). \\
%
%
In a more recent paper, Audit et al. (1997) have proposed some
analytic prescriptions to compute the collapse time along the
second and the third principal axes of an ellipsoid, by means of
the 'fuzzy'  threshold approach. They pointed out that the
formation of virialized clumps must correspond to the third
axis collapse and that the collapse along this axis is slowed down
by the  effect of the shear rather than be accelerated by it, in
contrast to its effect on the first axis collapse. They concluded
that spherical collapse is the fastest, in disagreement with
Bertschinger \& Jain's theorem.

Van de Weygaert \& Babul (1994) studied the influence of shear
fields on the evolution of galactic scale perturbations. They
found that the tidal forces induced by large-scale inhomogeneities
can affect the mass in and around primordial density peaks and
that in some peculiar conditions the shear can break up a
primordial peak into two distinct halos. Shear is even able to
produce the collapse of a void (van de Weygaert 1996).
%
%

Internal shear has been studied by means of collapsing homogeneous
ellipsoids (Icke 1973; WS; BS). These models show that the
evolution of an {\it isolated} homogeneous ellipsoid (namely an
ellipsoid not exposed to external tidal force) proceeds through a
series of uniform ellipsoids, or in other terms the shape of a
perturbation does not change much until it acquires a significant
overdensity with respect to the background. The collapse time of
perturbations of given initial overdensity, decreases with
increasing eccentricity and the collapse is faster for near-oblate
configurations (than near prolate ones). BS showed that the
density contrast at turn-around and the collapse velocity at
pancake formation are reduced relative to the spherical case.
Summarizing, the main conclusions are that internal shear can
alter the collapse history of structures (Icke 1973; WS). I want to
recall that the ellipsoid model has also been used to calculate
the cosmological density parameter (WS; BS; Watanabe 1993).

It is important to remark that while the homogeneous ellipsoid
model has been widely used in the cosmological context (Palmer
1983; WS; Peebles 1980; BS; Hoffman 1986; Monaco 1995; Eisenstein
\& Loeb 1995; Bond \& Myers 1996a,b; van de Weygaert 1996; Audit
\& Alimi 1996), only a few papers deal with the study of the effect
of shear on the ellipsoid evolution and
the analytical studies are even less
(WS; BS).
Moreover, the
effect of internal and external shear has been studied separately:
for example WS, BS and Watanabe \& Inagaki (1991, 1992) 
neglect the role of the tidal forces; Watanabe 1993 studied only the
effect of external shear.

It is then interesting to study the evolution of the
homogeneous ellipsoid taking account of both internal and
external shear and to look for analytical solutions of the
equations describing the evolution of an ellipsoidal perturbation.

In this paper, I shall study the effect of internal and external
shear finding an approximate analytical solution to the equations
of motion given in Watanabe (1993). These equations are the
equations of Icke (1973) and WS, modified to take account of the
effect of the tidal distortion. Similarly to WS, I
find an approximate solution to Watanabe's equations which is
after compared with numerical simulations.

The paper has fundamentally two aims:\\
1) to find an analytical solution for the equations describing the
evolution of an ellipsoidal perturbation and to study the changes
in configuration, axial velocity fields, and density contrast in
terms of the internal shear (intrinsic initial asphericity) and
external shear.  \\
2) To study how internal and external shear affect the ellipsoid
overdensity at turn-around and axial velocity at the collapse
epoch. The final aim is to have some insights in the
previrialization conjecture.

I develop this last item by improving BS model, namely I
use the analytical solution of the equations of motion, previously
found, to calculate the density contrast at turn-around and the
axial velocity at the collapse epoch, similarly to what is done by BS.

The plan of the paper is the following: in Sect. ~2, I
introduce the equations of motion of an {\it unisolated}
ellipsoid. In Sect. ~3, I find an approximate analytical
solution to the previous equations, similarly to what is done by WS
for the {\it isolated} ellipsoid model. In Sect. ~4, I use the
previous solution to find the evolution of the density contrast and
the axial peculiar velocity. In Sect. ~5, the density contrast
at turn-around and the velocity at collapse are calculated in order
to have some insights on the effects of shear on the collapse of
the structure. Sect. ~6 is devoted to conclusions.

\section{Unisolated homogeneous ellipsoid model}

The equations of motion of an irrotational homogeneous ellipsoid
with semiaxis lengths $a_1$, $a_2$ and $a_3$ embedded in a
Friedmann-Robertson-Walker background was given by Icke (1973), and WS
as:

\begin{equation}
\frac{d^2a_{\rm i}}{dt^2}=-2\pi G\left[ \rho _{\rm e}\alpha _{\rm
i}+\left( \frac 23-\alpha _{\rm i}\right) \rho _{\rm b}\right]
a_{\rm i}
 \label{eq:WS}
\end{equation}
where $R_{\rm b}$ and $\rho_{\rm b}$ are respectively the scale
factor and density of the background universe, $\rho_{\rm e}$ the
density within the ellipsoid. The coefficients $\alpha_{\rm i}$
are given by:
\begin{equation}
\alpha _{\rm i}=a_1a_2a_3\int_0^\infty \frac{d\lambda }{\left(
a_{\rm i}^2+\lambda \right) \left[ \left( a_1^2+\lambda \right)
\left( a_2^2+\lambda \right) \left( a_3^2+\lambda \right) \right]
^{\frac 12}} \label{eq:alpi}
\end{equation}
For the rest of the paper, except section~(5), I assume that $a_1
\leq a_2 \leq a_3$. Note that $\alpha_{\rm i}$ satisfies
$\alpha_1+\alpha_2+\alpha_3=2$, condition coming from the Poisson
equation. The equations to solve are then Eq. ~(\ref{eq:WS})
together with Eq. ~ (\ref{eq:alpi}) and the cosmological
equation for the Friedmann model:
\begin{equation}
\frac{d^2R_{\rm b}}{dt^2}=-\frac{4\pi }3G\rho _{\rm b}R_{\rm b}
\label{eq:Fried}
\end{equation}
where $R_{\rm b}$ is the scale factor of the background universe,
and the equations of the conservation of mass in each region:
\begin{equation}
\rho_{\rm e} a_1a_2a_3={\rm const} \hspace{0.5cm} \rho_{\rm
b}a_1a_2a_3={\rm const} \label{eq:dens}
\end{equation}
Eqs. ~(\ref{eq:WS})-(\ref{eq:dens})
describe the evolution of an ${\it isolated}$ ellipsoid.

Before going further, let us consider the effects of the
over-simplifications of the model:\\
1) The ellipsoidal approximation, described by the 
Eqs. (\ref{eq:WS})-(\ref{eq:dens}), applies mainly in the
immediate vicinity of the density
extrema (maxima and minima), where the leading terms in the
gravitational potential are the quadratic ones. Previous works (see Eisenstein \& Loeb 1995) has
shown that the inner region of a perturbation is well approximated by the
ellipsoid approximation (Eisenstein \& Loeb 1995 apply this model to a
variety of mass scales in the range $10^8-10^{15} M_{\odot}$ and WS to superclusters.)\\
2) The homogeneous {\it isolated} ellipsoid model does not take
account of the role of the tidal forces due to other nearby objects. This problem can
be solved as shown in the following of this paper or as shown by Eisenstein \& Loeb (1995).\\
3) The homogeneous ellipsoid model does not take account of
the inhomogeneity and of the substructure internal to the system. This
last item is a natural limitation of the model: by definition, a
homogeneous ellipsoid cannot represent the substructure or
inhomogeneity of the object. This limitation, from one side,
prevents us from treating the distribution of matter and angular
momentum within the collapsing object and, from the other side, has the
effect of underestimating the effect of previrialization, and in
particular the value of the overdensity at virialization,
$\delta_{\rm v}$ (Peebles 1990). This means that the
effect of the shear on the evolution of the density contrast at
turnaround and the velocity at collapse, estimated with the ellipsoid model,
is even smaller than
what one expects in the collapse of a real protostructure.\\
%
%
4) Another limitation of the model is the effect of the matter in
the ellipsoid on the smooth background. One expects that the
ellipsoid matter causes the density of its immediate surroundings
to deviate from the cosmic mean. Then, as a back reaction, tidal
fields due to the perturbed external material should induce
departure from homogeneity in the ellipsoid. However, when the
density inside the ellipsoid, $\rho_{\rm e}$, is very close to the
background density, $\rho_{\rm b}$, this effect is negligibly small
(Watanabe \& Inagaki 1991; WS), while if $\rho_{\rm e}>>\rho_{\rm
b}$, the evolution will be determined by its self-gravity (WS).
Moreover, numerical simulations have shown that it is a good
approximation to ignore departures from homogeneity when one
calculates the evolution of the axis ratio (Hui \& Bertschinger
1995).
%
%

In spite of the uncertainties listed above, the ellipsoid model
give
a good approximation for the evolution of any object which
collapsed in a fairly homogeneous manner (WS; Eisenstein \& Loeb
1995).

While it is not possible to take account of inhomogeneity,
the homogeneous {\it isolated} ellipsoid model can be modified in
order to take account of external shear as done, for example, by
Watanabe (1993). In order to get the quoted goal, one has to get a
multipole expansion of the gravitational potential, $\phi$. The
gravitational potential at time $t$ and comoving coordinates
(${\bf x}$), due to the field outside a comoving radius X is given
by:
\begin{equation}
\phi ({\bf x,}t)=\sum_l\phi _l({\bf x},t)=G\rho _{\rm b}R^2_{\rm b
}\sum_lx^l\int_{x^{^{\prime }}>X}\frac{\delta ({\bf x}^{^{\prime
}},t)}{x^{^{\prime }l+1}}P_l(\mu ^{^{\prime }})d^3x^{\prime }
\end{equation}
where $P_l$ are the Legendre polynomials. In this paper, as I am
primarily concerned with shear, I focus on the quadrupole
($l=2$) terms. In fact, the monopole component, $l=0$, generates
no force, since the potential is spatially constant, while the
dipole component, $l=1$, induces only streaming motions that
cannot alter the shape or induce any rotation. As shown by
Eisenstein \& Loeb (1995), in the standard CDM scenario, the dipole
is generated at large scales, so the object I am studying and its
neighborhood move as bulk flow with the consequence that the
angular distribution of matter is very small, and then the dipole
terms can be ignored. The quadrupole term, $l=2$, is the first
term dealing with the effect of tidal distortion:
\begin{equation}
\phi_{2}({\bf x,}t)= \frac{3}{2}G \rho _{\rm b} R^2_{\rm b} Q_{\rm
ij} x_{\rm i } x_{\rm j} \label{eq:quad}
\end{equation}
where the quadrupole tensor, $Q_{\rm ij}$, is given by:
\begin{equation}
Q_{\rm ij}=\int \frac{\delta \left( {\bf x}^{^{\prime }},t\right)
}{x^{^{\prime }5}}\left[ x_i^{^{\prime }}x_j^{^{\prime }}-\frac
13\delta _{ij}{\bf x}^{^{\prime }2}\right] d^3x^{^{\prime }}
\label{}
\end{equation}

The quadrupole tensor, $Q_{\rm ij}$, is a traceless $3 \times 3$
matrix, that can be diagonalized using an appropriate coordinate
transformation to get:
\begin{equation}
Q_{\rm ij}=Q \left(
\begin{array}{ccc}
-\beta & 0 & 0 \\
0 & \beta-1  & 0 \\
0 & 0 & 1
\end{array}
\right), \hspace{0.5cm}  1/2 \leq \beta \leq 1
 \label{}
\end{equation}
In the following, I assume that the field does not vary
significantly during the ellipsoid evolution and consequently, for
simplicity, the value of $\beta$ is assumed to be constant in time
(Watanabe 1993; van de Weygaert 1996). For a positive value of Q,
which is the component with maximum absolute value in the
quadrupole tensor, a perturbation evolves to a prolate
configuration and to an oblate one for a negative value of Q. If
the initial perturbation is spherically symmetric, it shall evolve
to an ellipsoid, if all the three components of the quadrupole
tensor are different, while, if two components of the quadrupole
tensor are equal and $Q>0$ the resulting figure is a prolate
spheroid or an oblate one if $Q<0$. 
If the initial perturbation is
an ellipsoid, also in the case that the quadrupole tensor has two
equal components, the final configuration is an ellipsoid.
In fact, though the cluster becomes a spheroid if the two axes of the external perturbations
are equal in Watanabe's (1993) model (because he assumed the initial
configuration to be spherical symmetric) generally, the external
perturbations are triaxial so the cluster remains a figure of ellipsoid
if the initial directions of axes of the cluster coincides with those of
external perturbations.

The equations of motion are obtained by adding the force due to
the potential given by Eq. ~(\ref{eq:quad}) into
Eq. ~(\ref{eq:WS}). Assuming
that the principal axes of the external tidal tensor are always
oriented along the principal axes of the mass tensor,
the evolution equations reduces to three equations for the three
semiaxes of the ellipsoid and are given by (Watanabe 1993;
van de Weygaert 1996):
\begin{eqnarray}
\frac{d^2a_{\rm i}}{dt^2}&=&-2\pi G\left\{ \rho _{\rm e}(\alpha
_{\rm i}-\gamma b_{\rm i})+\left[ \frac 23-(\alpha _{\rm i}-\gamma
b_{\rm i})\right] \rho _{\rm b}\right\} a_{\rm i} = \nonumber \\
& & -2\pi G\left[
\rho _{\rm e}\alpha _{\rm i}+\left( \frac 23-\alpha _{\rm
i}\right) \rho _{\rm b}\right] a_{\rm i}-2 \pi G \gamma
\left(-b_{\rm i}\right) \left(\rho_{\rm e}-\rho_{\rm b}
\right)a_{\rm i}
\label{eq:WSM}
\end{eqnarray}
where:
\begin{equation}
\gamma =\frac 3{2\pi }\frac Q\delta, \hspace{0.5cm} {\bf
b}=(-\beta ,\beta -1,1) \label{eq:bb}
\end{equation}
Note that in the rightmost term of Eq. ~(\ref{eq:WSM}), I wrote
$-b_{\rm i}$ to display the equation in the same form of van de
Weygaert (1996), in order to simplify the comparison with the
result of that paper. A comparison of Eq. ~(\ref{eq:WS}) with
Eq. ~(\ref{eq:WSM}) shows that the evolution of the ellipsoid
is modified by the tidal force if
the term $\gamma b_{\rm i}$ is large, while the evolution of the
ellipsoid is dominated by self-gravity if this term is small. I
implicitly assume that the external structures, giving rise to the
tidal field, are at large distance from the ellipsoid (see
Eisenstein \& Loeb 1995). As a consequence, the amplitude of the
external quadrupole force is assumed to increase with the linear
growth rate (Ryden 1988; Watanabe 1993; Eisenstein \& Loeb 1995),
$D(t)$ (this last quantity is given in Peebles 1980):
\begin{equation}
Q(t)=Q_0 \frac{D(t)}{D_0}
\label{eq:quadru}
\end{equation}
Here the subscript ``0" means that the corresponding quantity is
calculated at the present epoch and $D(t)=R_{\rm b}(t)$, for an
Einstein-de Sitter (hereafter EdS) universe.
Similarly to WS, I assume that initially $R_{\rm b}=1$. In the particular case of an {\it isolated} ellispoid with
$a_1:a_2:a_3=1:1.25:1.5$, $R_{\rm b}=873$ at collapse. For sake of precision, I want to stress that in an Einstein-de Sitter universe, Eq. (\ref{eq:quadru}), 
$Q(t)=Q_0 \frac{R_{\rm b}(t_0) (t/t_0)^{2/3}}{R_{\rm b}(t_0)}$ reduces to 
$Q(t_0)=Q_0 $
at present time, $t_0$.

In order to have an estimate of the value of $Q_0$,
for a cluster interacting with a neighboring one, I use the simple
model in Watanabe (1993). I consider a cluster which has a
neighboring cluster with a mean density contrast $<\delta> \simeq
3$, a comoving separation $(0,0,x_3)$, and a comoving size $\Delta
x_3=x_3/3$. The $Q_{33}$ quadrupole component is given by:
\begin{equation}
Q_{33} \simeq \frac{8}{9} \pi <\delta> \left(\frac{\Delta
x_3}{x_3}\right)^3 \simeq 0.3 \label{}
\end{equation}
The previous estimate corresponds to a cluster interacting with a
neighbor having mass excess comparable to that of the Virgo
cluster, and separation three times its size.

Another way of estimating $Q_0$ is by using the anisotropy of the velocity field in the 
LSC from data of Lilje, Yahil \& Jones (1986). If we indicate with $Q_{\rm v0}$ the component of the 
largest absolute value of the anisotropic velocity, one gets:
$Q_0 \Omega_0^{0.6}=\frac{4 \pi}{3} Q_{\rm v0}$ (Watanabe 1993). Since Lilje, Yahil \& Jones (1986) deduced a value of 
$Q_{\rm v0} \sim 0.1-0.2$ at the distance of the Local Group from Virgo, we have that
$Q_0 \Omega_0^{0.6} \sim 0.4-0.8$. 

Before going on, it is important to discuss a basic difference between the present paper 
and that of Watanabe (1993). 
Differently from Watanabe (1993), in this paper I assume 
that protostructures have an initial asphericity, 
while the paper of Watanabe (1993) (similarly to those of van de Weigaert (1996) and Palmer (1983)) assume that 
the initial configuration is spherical, so that the principal axes of the external tidal tensor 
will be oriented along the principal axes of the mass tensor and the equations of motions 
reduces to three equations involving the diagonal components of the traceless tidal tensor. 
Our assumption of initial asphericity of protostructures, is motivated by the fact that previous analyzes of the topology 
of the constant-density profiles in the neighborhood of the peaks of the Gaussian field showed that 
the isodensity surfaces 
are simply connected and approximately ellipsoidal 
(Doroshkevich 1970; Bardeen 1986).
We also know that 
the initial asphericity has a certain role in shaping the final configuration 
of the structure (Icke 1973; WS; BS). By means of the assumption,
we have the noteworthy advantage of studying the joint effect of ``internal and external shear"
(see the final part of Sect. ~ (3) for a discussion).
The approach of this paper, 
assuming that the principal axes of the external tidal tensor are always
oriented along the principal axes of the mass tensor, is dictated by 
reasons of mathematical simplicity.
%
%
I must tell that, at the same time, the assumption is not strange or without motivation: for example 
van de Weygaert \& Babul 1994, in order to study the effect of shear fields on the evolution of galactic scale density peaks, do a similar assumption, namely that the shear tensor, at the location of the peak representing the structure, is oriented so that it is diagonal. Moreover in a recent paper Porciani, Dekel \& Hoffman (2002) find a stronly alignement 
between the principal axes of the inertia and shear tensor, in contraddiction to usual assumption that the two tensors 
are largely uncorrelated (Hoffman 1986b; Heavens \& Peacock 1988; Catelan \& Theuns 1996).
%
%

Finally, I want to stress that the approximate solution found in the present paper give a more general representation 
of structure formation than those described, as examples, in the following sections. For example, the assumptions 
and results of Watanabe (1993) are re-obtained assuming that the three axes of the ellipsoid are equal, or in other words 
Watanabe's result is a ``particular case" of those of this paper when $a_1(t_{\rm i})=a_2(t_{\rm i})=a_3(t_{\rm i})$. Moreover, in the present paper, I get analitycal solutions for the equations of motion of the ellipsoid, while 
Watanabe (1993) solve the same equations numerically.

\section{An analytical approximate solution}

An analytical approximation of the solution describing the
evolution of the $i-th$ axis of the {\it isolated} ellipsoid was
found by WS. The equations of motion can be integrated
analytically if one assumes that:\\
1) The evolution of the configuration is self-similar, which means
that $\alpha_{\rm i}(t)\simeq \alpha_{\rm i}(t_{\rm i})$.\\
2) The time dependence of $\rho_{\rm e}a_{\rm i}$ and $\rho_{\rm
b} a_{\rm i}$ are the same as the spherical model, namely:
\begin{equation}
(\rho_{\rm e}a_{\rm i})(t)=\frac{-3}{4 \pi G} \ddot R_{\rm s }
\frac{R_{\rm b}(t_{\rm i})}{R_{\rm s}(t_{\rm i})}= \frac{-3}{4 \pi
G} \ddot R_{\rm e } \frac{R_{\rm b}(t_{\rm i})}{R_{\rm e}(t_{\rm
i})} \label{eq:simp}
\end{equation}
where $R_{\rm s }$ is the radius of a spherical shell whose
initial density enhancement within $a_{\rm s }$ is $\delta_{\rm
s}(t_{\rm i})=\delta_{\rm e}(t_{\rm i})$ and
\begin{equation}
(\rho_{\rm b}a_{\rm i})(t)=\frac{-3}{4 \pi G} \ddot R_{\rm b}
\frac{a_{\rm i}(t_{\rm i})}{R_{\rm b}(t_{\rm i})} \label{eq:simp1}
\end{equation}
Defining $R_{\rm s}(t_{\rm i})=R_{\rm b}(t_{\rm i})=1$, and
substituting Eqs.(\ref{eq:simp})-(\ref{eq:simp1}) into
Eq. (\ref{eq:WS}) and integrating, one obtains:
\begin{equation}
\frac{a_{\rm i}(t)}{a_{\rm i}(t_i)}=R_{\rm b}-\frac 32\alpha _{\rm
i}\left( R_{\rm b}-R_{\rm e}\right) \label{eq:pred}
\end{equation}
(WS). This approximation gives a good representation for the
evolution of the semiaxes of the ellipsoid for configurations not
too extreme. A comparison between the numerical solution of the
equations of motion, obtained using a Bulirsch-Stoer scheme, and
the prediction of Eq.~(\ref{eq:pred}) is shown in Fig. 1a-1c.
In all the cases studied, the ellipsoids are embedded in an
EdS background universe with Hubble
parameter $H_0=50 {\rm km/s/Mpc}$ and $\rho_{\rm e}/\rho_{\rm
b}=1.003$. The velocity perturbation is taken to correspond to the
growing mode solution of the linear perturbation theory.
Calculations are terminated when the shortest axis becomes zero. In
\begin{figure}
\label{Fig. 1} \centerline{\hbox{Fig. 1 (a)
\psfig{figure=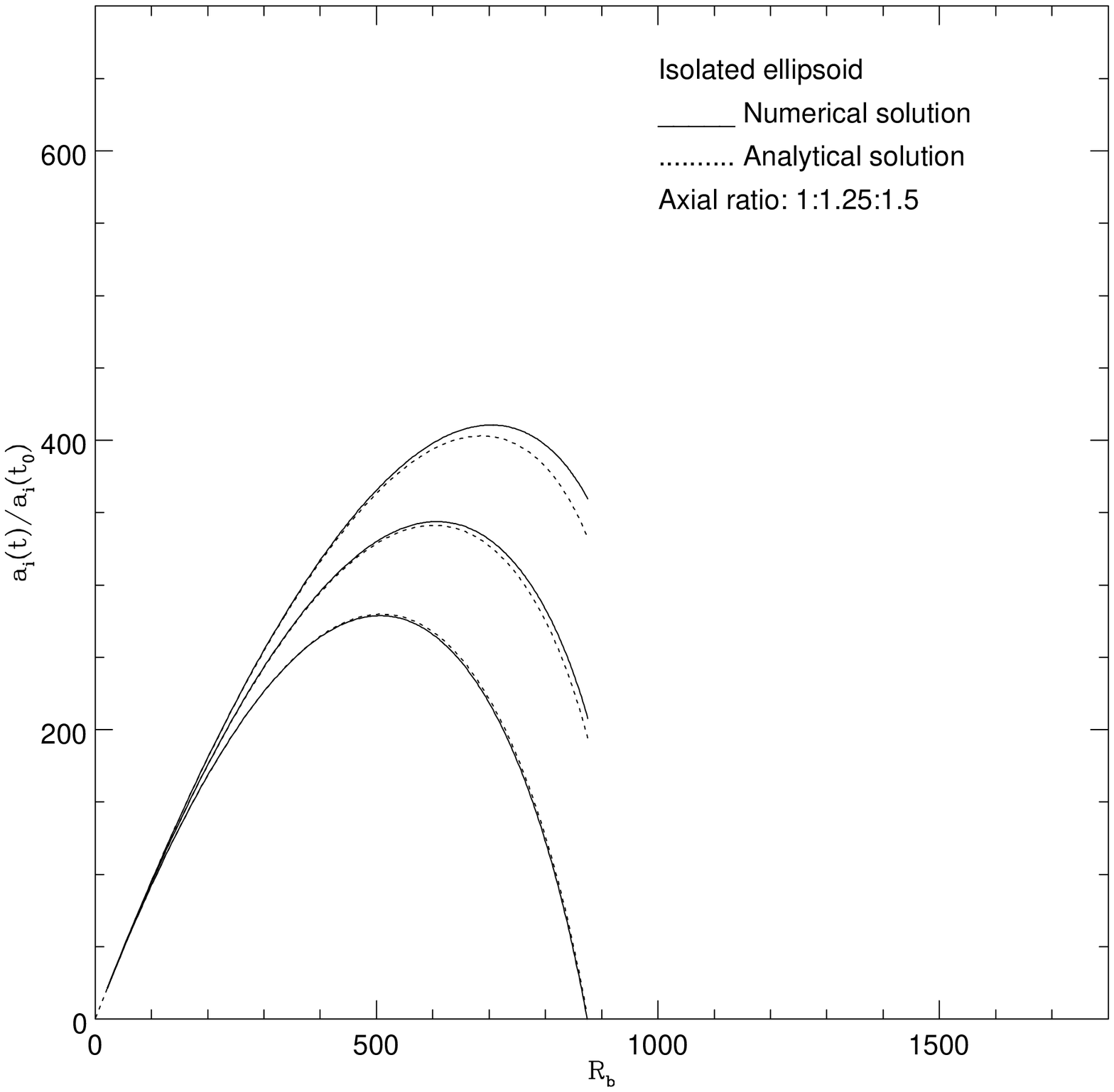,width=18cm}
}}
{\bf Fig. 1a} Evolution of {\it isolated} homogeneous ellipsoidal
perturbations in an EdS universe with $H_0=50 {\rm km/s/Mpc}$, 
$\rho_{\rm e}/\rho_{\rm b}=1.003$ and axial ratio $1:1.25:1.5$. 
The solid lines represent
numerical solutions obtained using Bulirsh-Stoer algorithm while
the dotted ones the approximate analytical solution of WS. 
\end{figure}

\begin{figure}
\label{Fig. 1} \centerline{\hbox{
Fig. 1 (b)
\psfig{figure=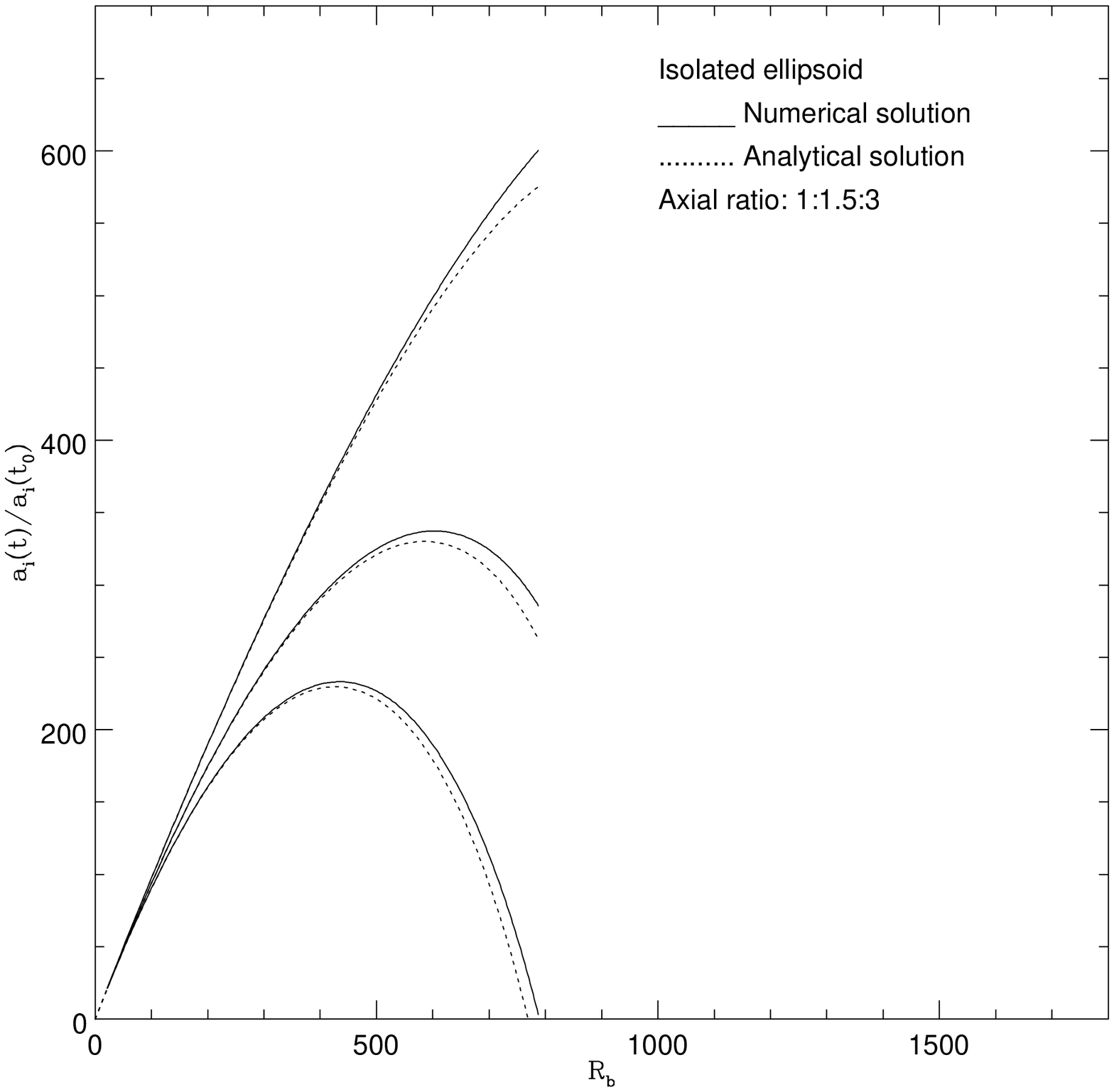,width=18cm}
}}
 {\bf Fig. 1b} Same as Fig. 1a but now the axial ratio is $1:1.5:3$.
\end{figure}

\begin{figure}
\label{Fig. 1} \centerline{\hbox{
Fig. 1 (c)
\psfig{figure=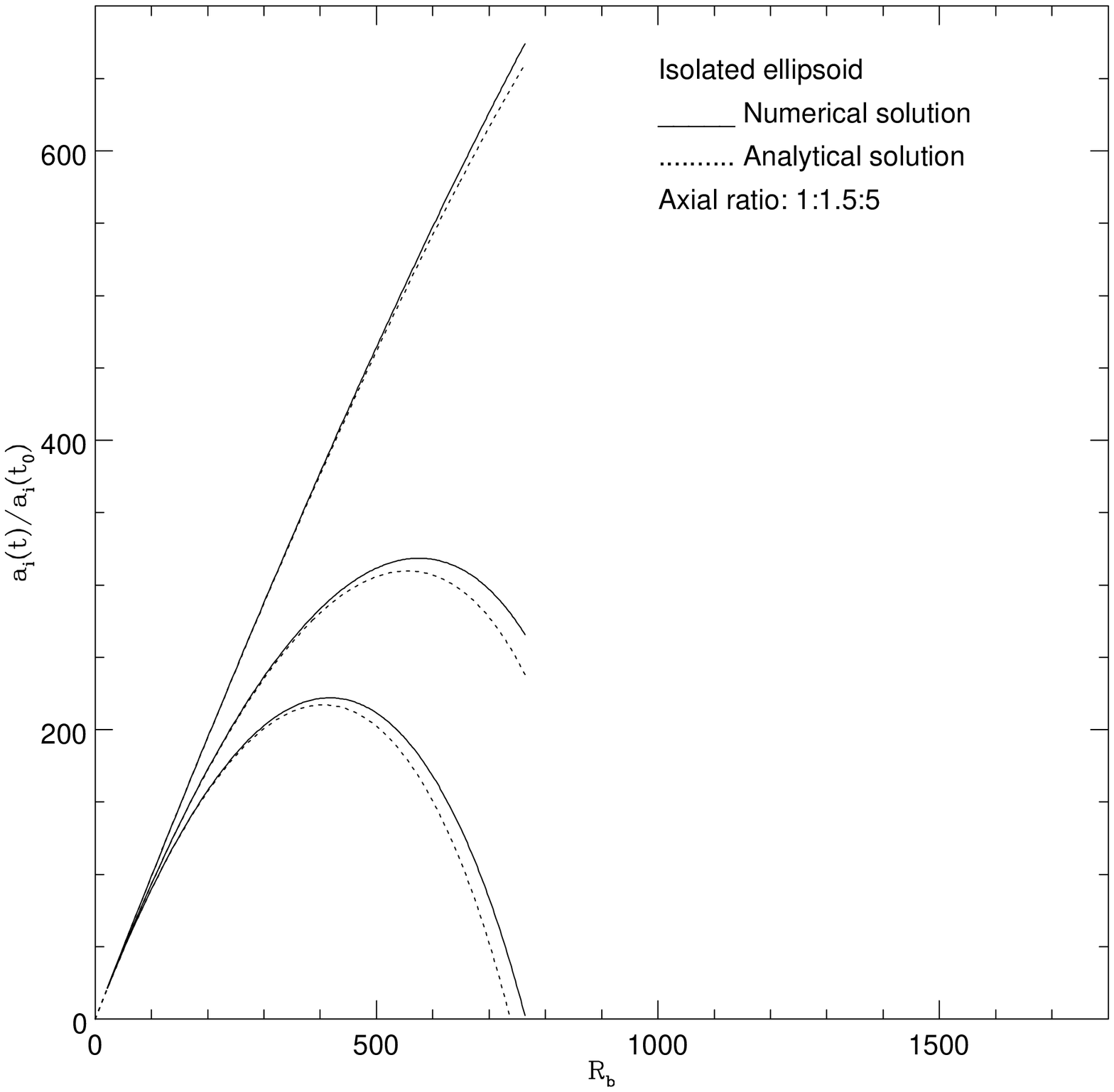,width=18cm}  }}
 {\bf Fig. 1c} Same as Fig. 1a but now the axial ratio is $1:1.5:5$.
\end{figure}

\begin{figure}
\label{Fig. 1} \centerline{\hbox{
Fig. 1 (d)
\psfig{figure=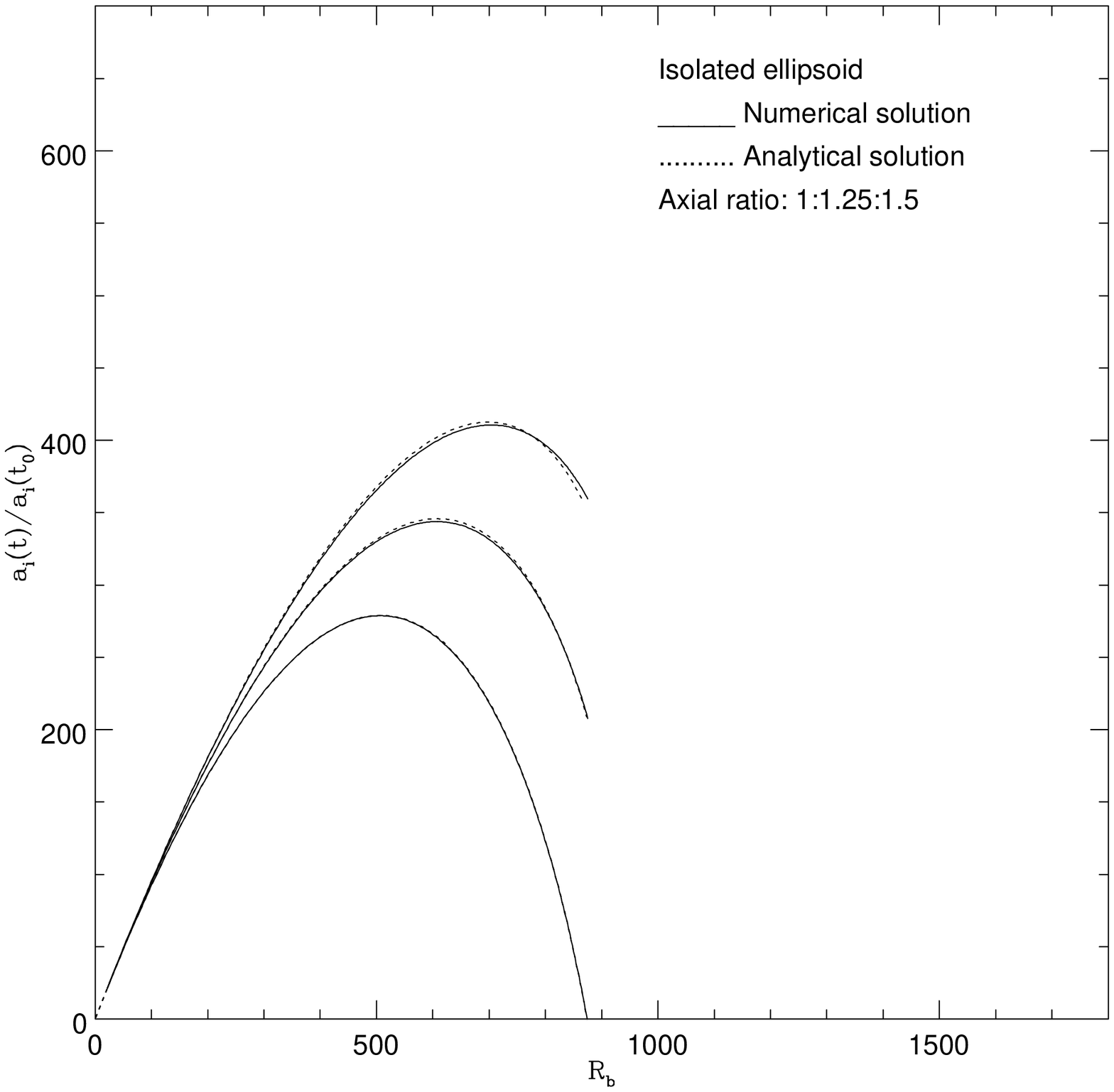,width=18cm}  }}
 {\bf Fig. 1d} Same as the previous figure
Fig. 1a but now the WS analytical approximate solution is
improved by using some free parameters fitted to the numerical
solution by means of the least-square method (see text for a
description).
\end{figure}

\begin{figure}
\label{Fig. 1} \centerline{\hbox{
Fig. 1 (e)
\psfig{figure=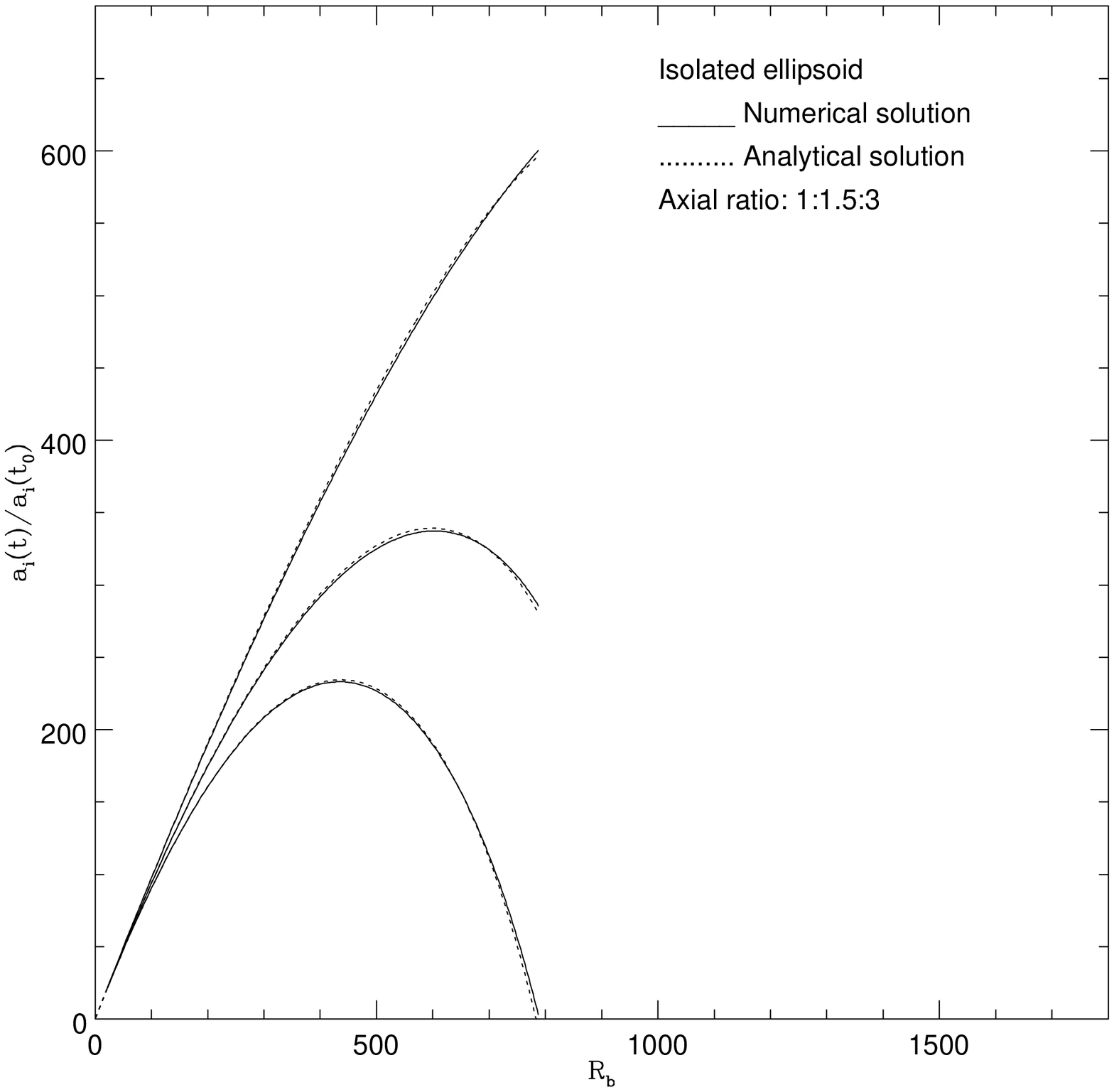,width=18cm}  }}
 {\bf Fig. 1e} Same as the previous figure
Fig. 1b but now the WS analytical approximate solution is
improved by using some free parameters fitted to the numerical
solution by means of the least-square method (see text for a
description).
\end{figure}

\begin{figure}
\label{Fig. 1} \centerline{\hbox{
Fig. 1 (f)
\psfig{figure=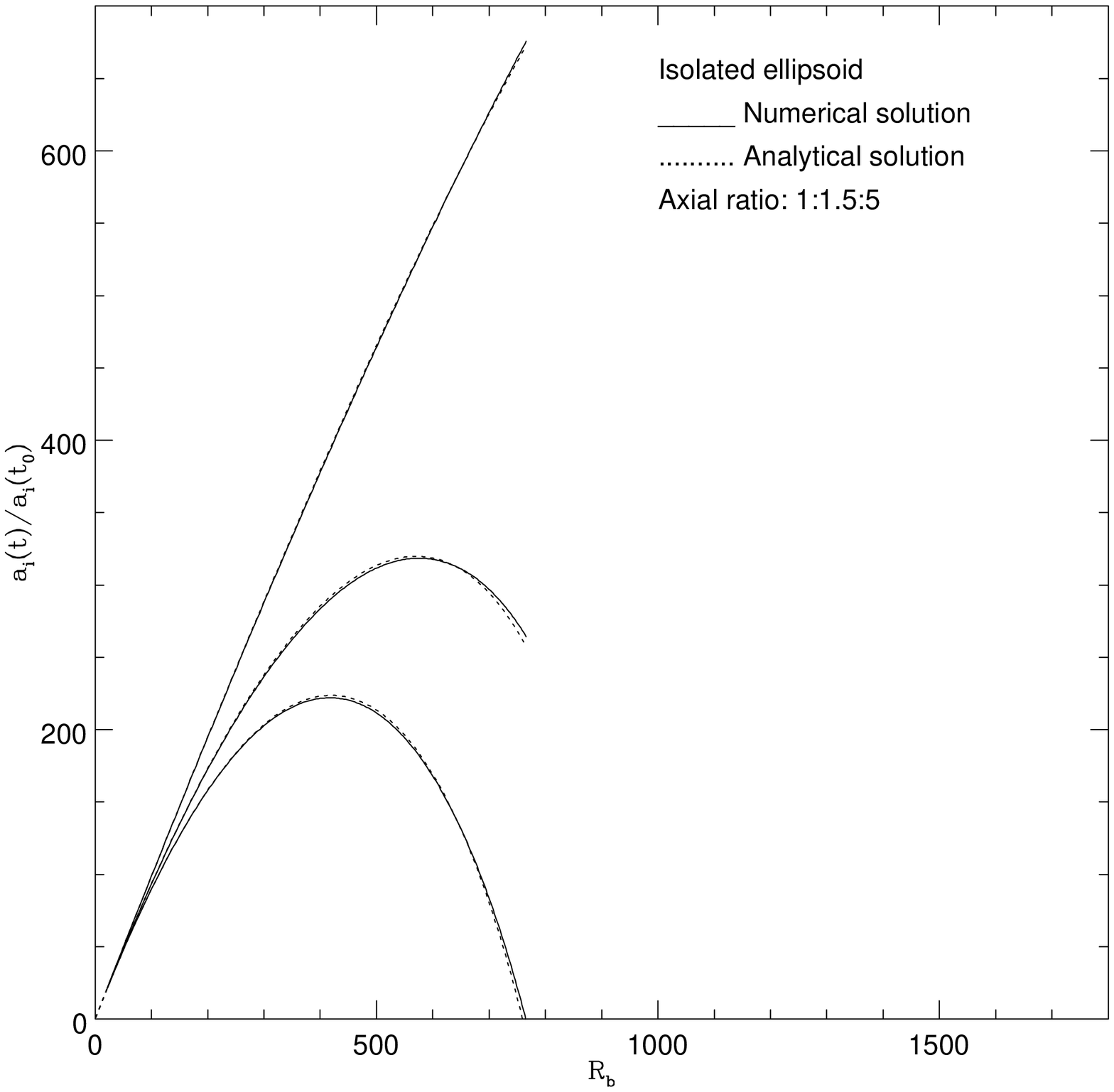,width=18cm}  }}
 {\bf Fig. 1f} Same as the previous figure
Fig. 1c but now the WS analytical approximate solution is
improved by using some free parameters fitted to the numerical
solution by means of the least-square method (see text for a
description).
\end{figure}

\begin{figure}
\label{Fig. 2} \centerline{\hbox{
Fig. 2 (a)
\psfig{figure=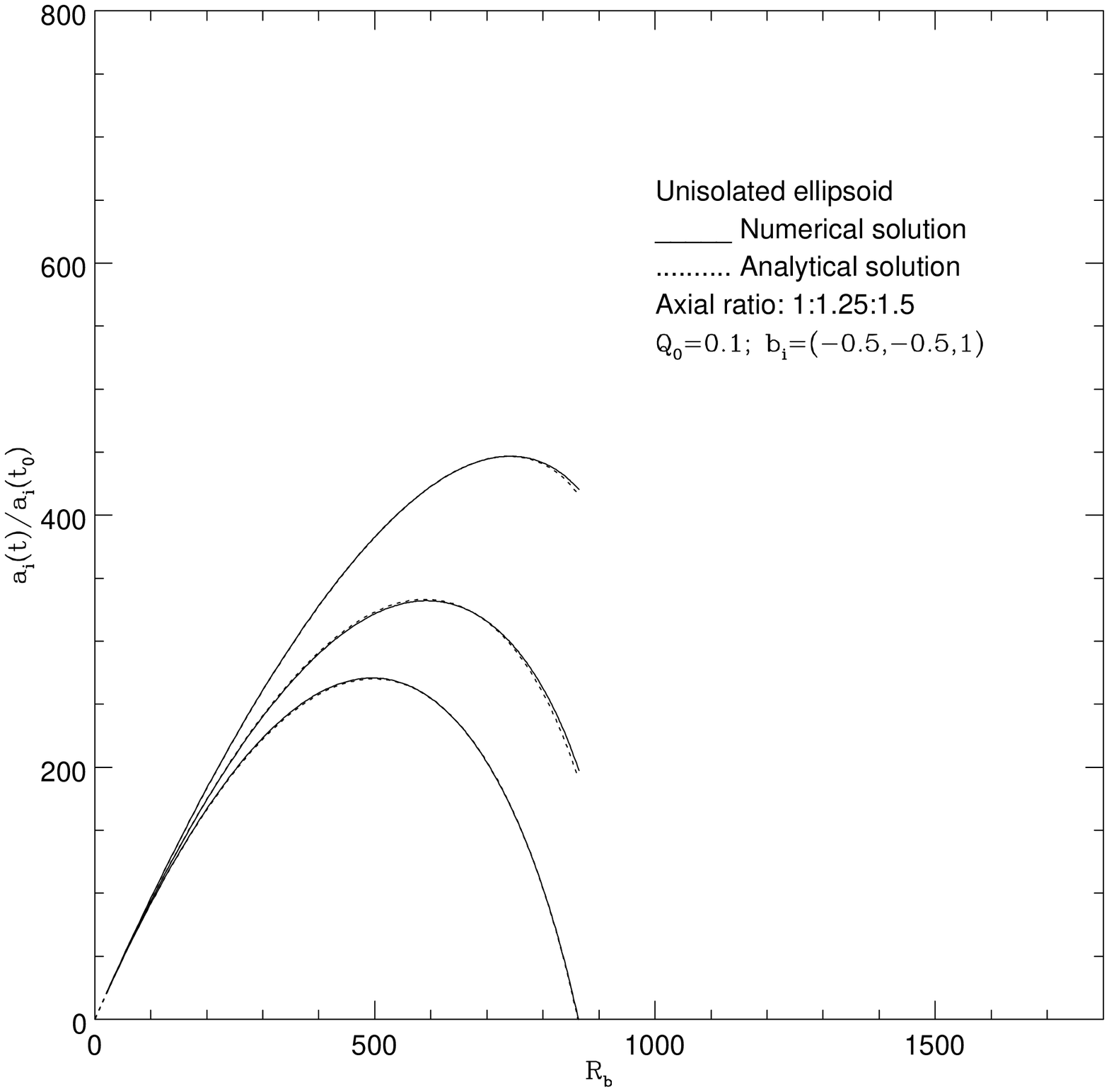,width=18cm}  }}
 {\bf Fig. 2a} Evolution of {\it unisolated} homogeneous ellipsoidal
perturbations in an EdS universe with $H_0=50 {\rm km/s/Mpc}$, 
$\rho_{\rm e}/\rho_{\rm b}=1.003$, axial ratio is $1:1.25:1.5$, 
$b_{\rm i}=(-0.5,-0.5,1)$, and $Q_0=0.1$. 
\end{figure}

\begin{figure}
\label{Fig. 2} \centerline{\hbox{
Fig. 2 (b)
\psfig{figure=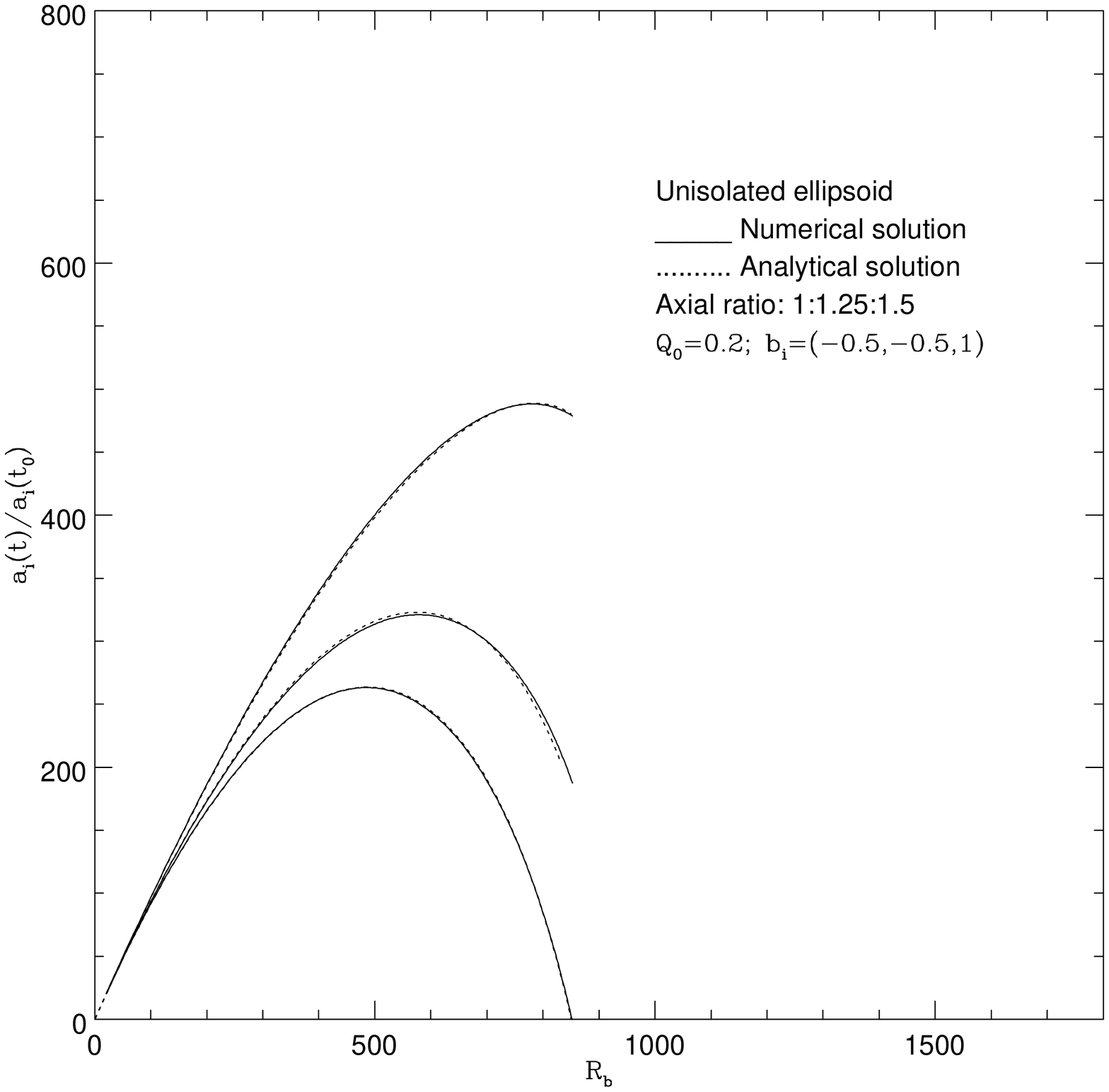,width=18cm}  }}
 {\bf Fig. 2b} Same as Fig. 2a but now $Q_0=0.2$
\end{figure}
\begin{figure}
\label{Fig. 2} \centerline{\hbox{
Fig. 2 (c)
\psfig{figure=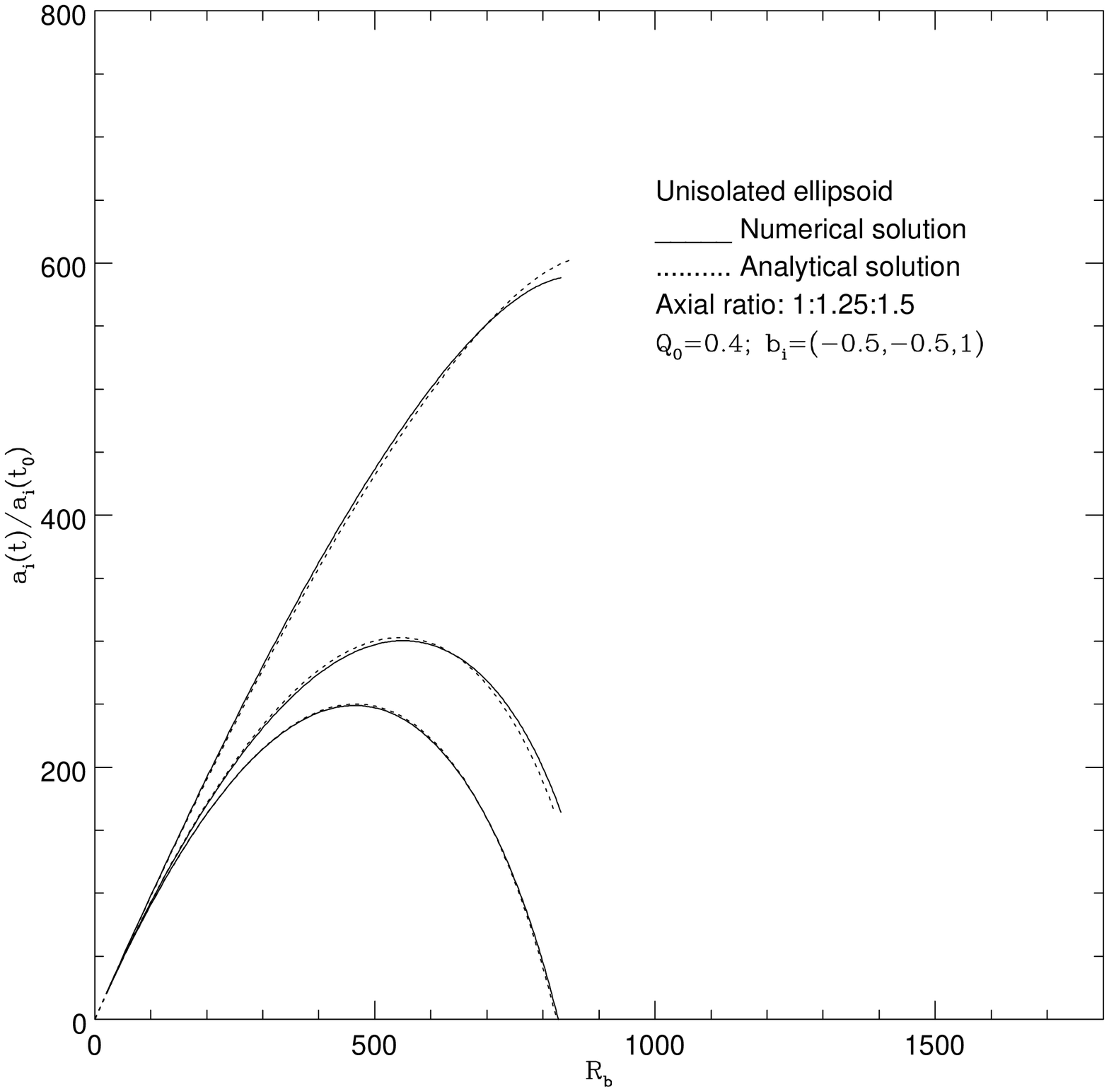,width=18cm}  }}
 {\bf Fig. 2c} Same as Fig. 2a but now $Q_0=0.4$
\end{figure}
\begin{figure}
\label{Fig. 2} \centerline{\hbox{
Fig. 2 (d)
\psfig{figure=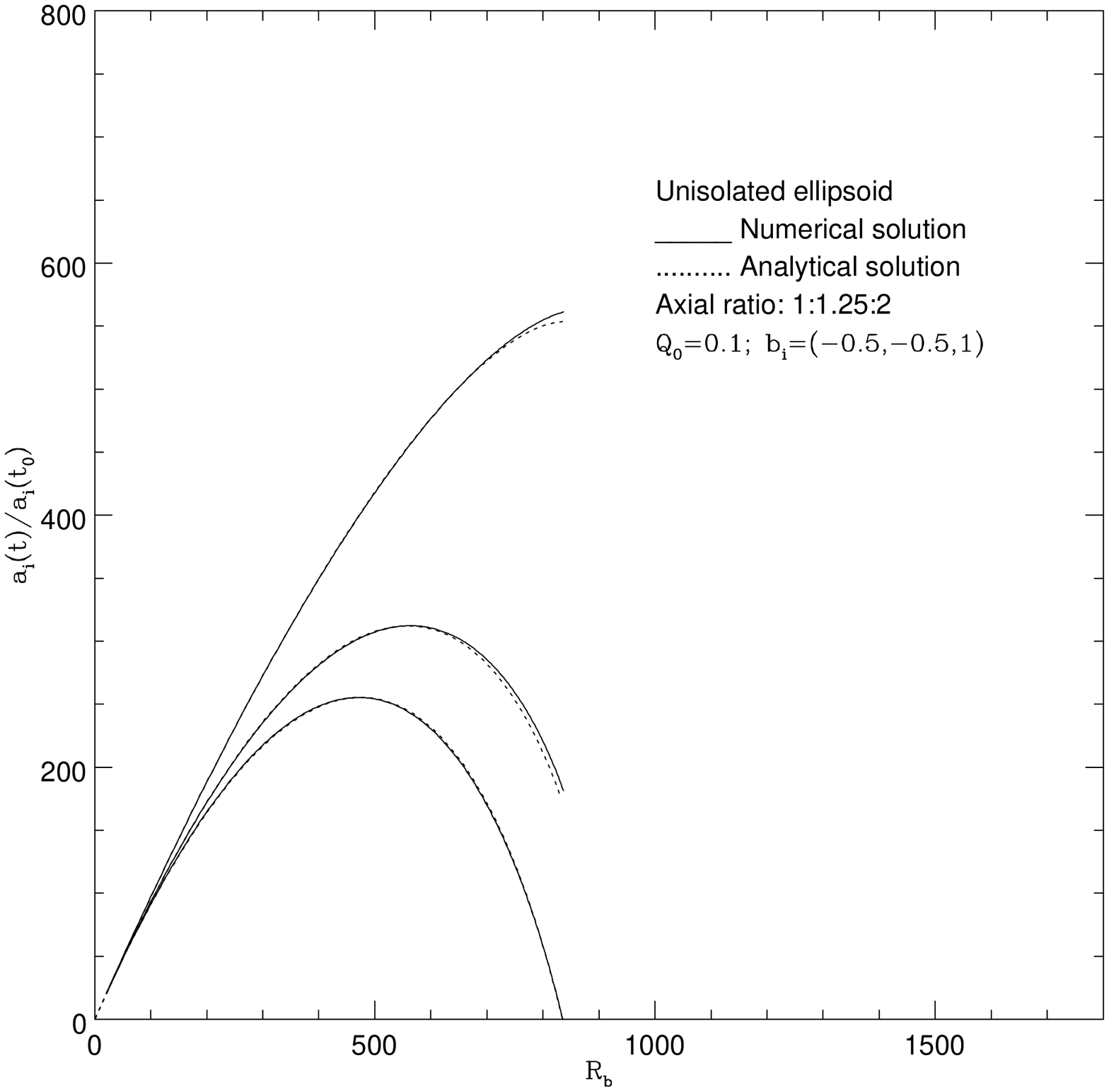,width=18cm}  }}
 {\bf Fig. 2d} Same as the previous
figure 2a but now the axial ratio is 
$1:1.25:2$, while $Q_0=0.1$ and $b_{\rm i}=(-0.5,-0.5,1)$.
\end{figure}
\begin{figure}
\label{Fig. 2} \centerline{\hbox{
Fig. 2 (e)
\psfig{figure=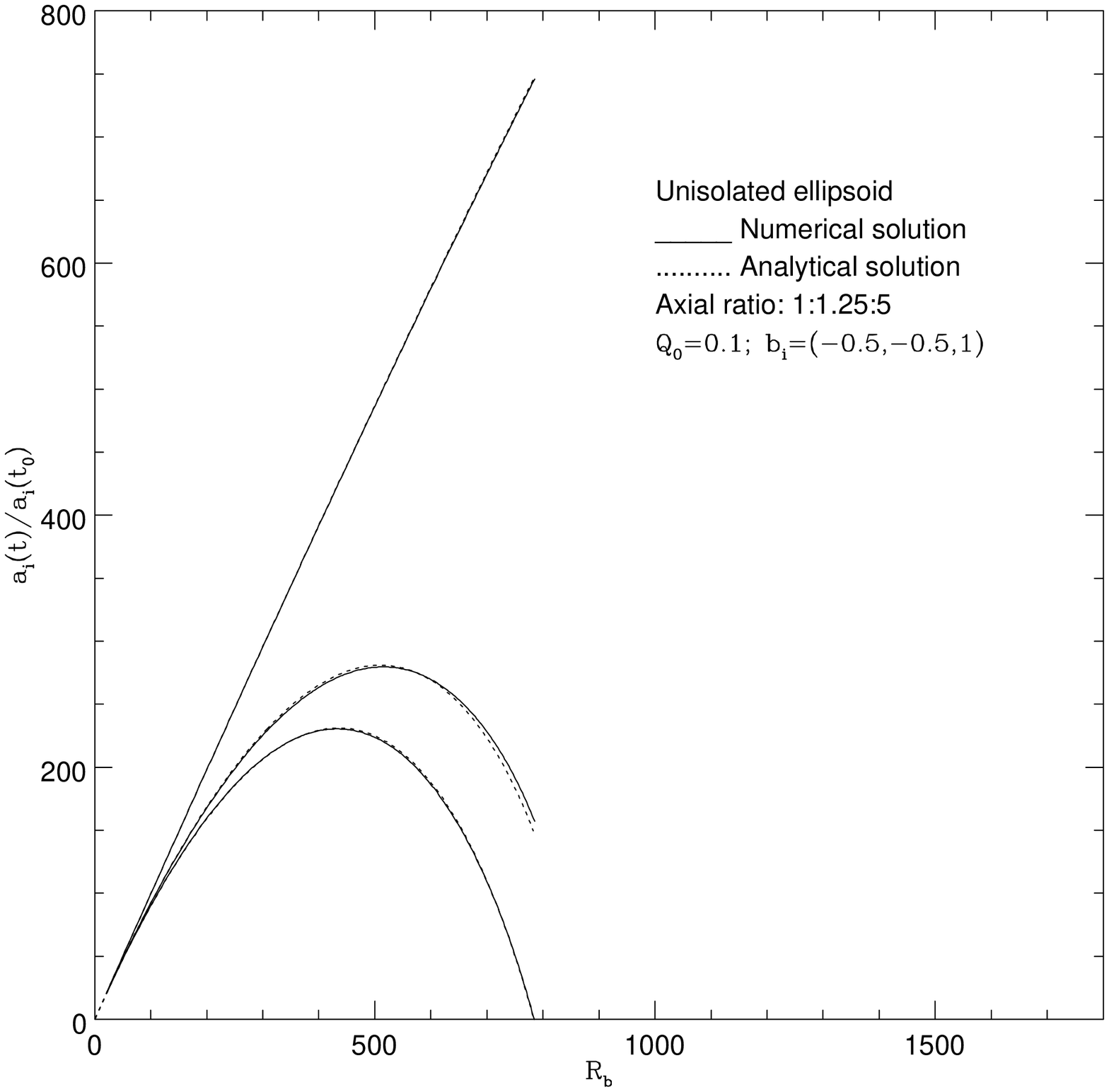,width=18cm}  }}
 {\bf Fig. 2e} Same as the previous
figure 2a but now the axial ratio is 
$1:1.25:5$, while $Q_0=0.1$ and $b_{\rm i}=(-0.5,-0.5,1)$.
\end{figure}

\begin{figure}
\label{Fig. 3} \centerline{\hbox{
Fig. 3 (a)
\psfig{figure=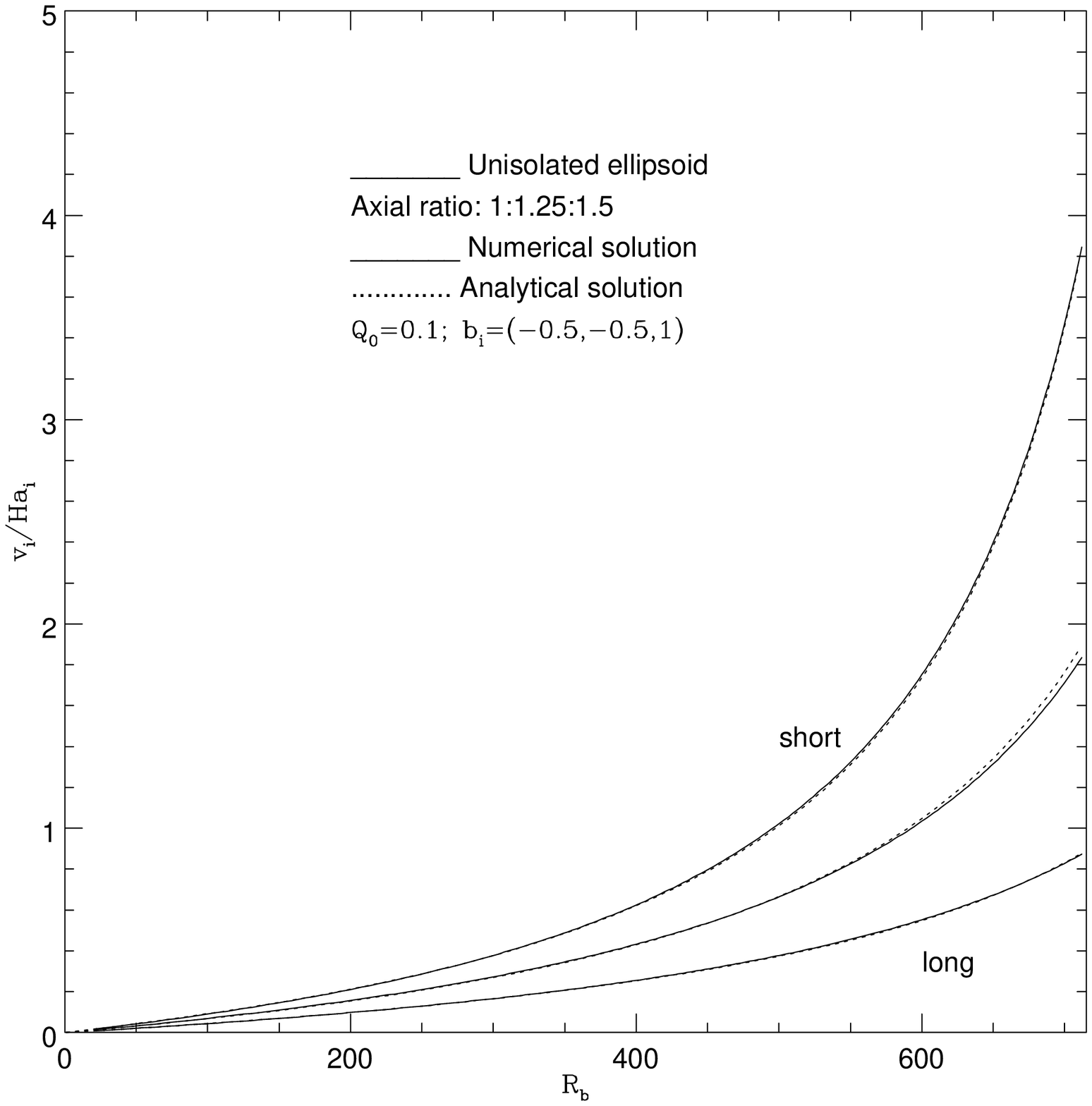,width=18cm}  }}
 {\bf Fig. 3a} Evolution of the axial velocity. The solid
lines represent, from up to bottom, the numerical result for the
peculiar velocity along the shortest, medium and longest axis. The
dotted line represents the velocity obtained from Eq.
(31) using the approximation for the semiaxes (Eqs.
(22)-(24)). The value of the axial ratio is $1:1.25:1.5$, while $Q_0=0.1$, and $b_{\rm
i}=(-0.5,-0.5,1)$. 
\end{figure}
\begin{figure}
\label{Fig. 3} \centerline{\hbox{
Fig. 3 (b)
\psfig{figure=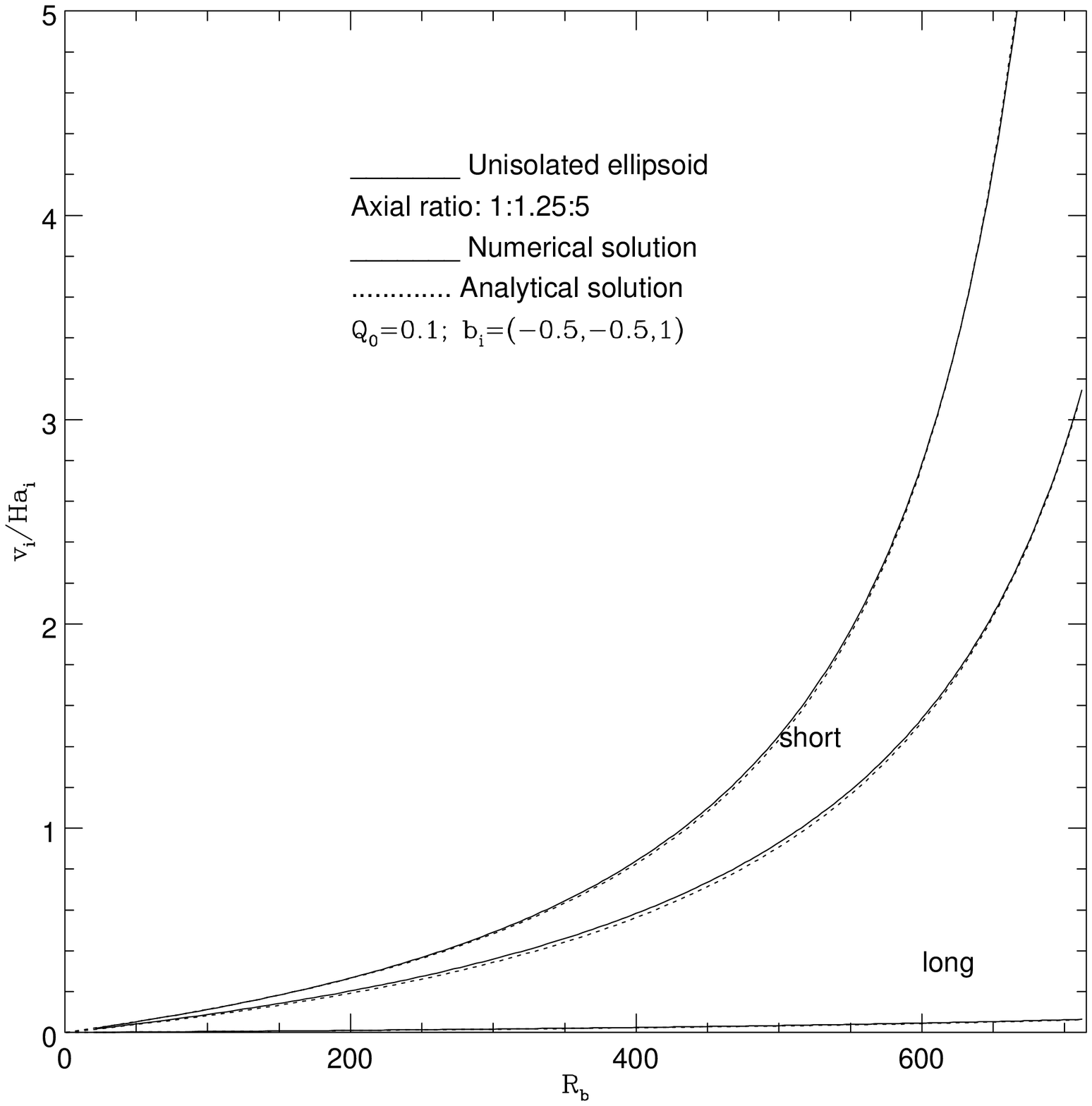,width=18cm}  }}
 {\bf Fig. 3b} Same as Fig. 3a but now the value of the axial ratio is $1:1.25:5$, while $Q_0=0.1$, and $b_{\rm
i}=(-0.5,-0.5,1)$. 
\end{figure}
\begin{figure}
\label{Fig. 3} \centerline{\hbox{
Fig. 3 (c)
\psfig{figure=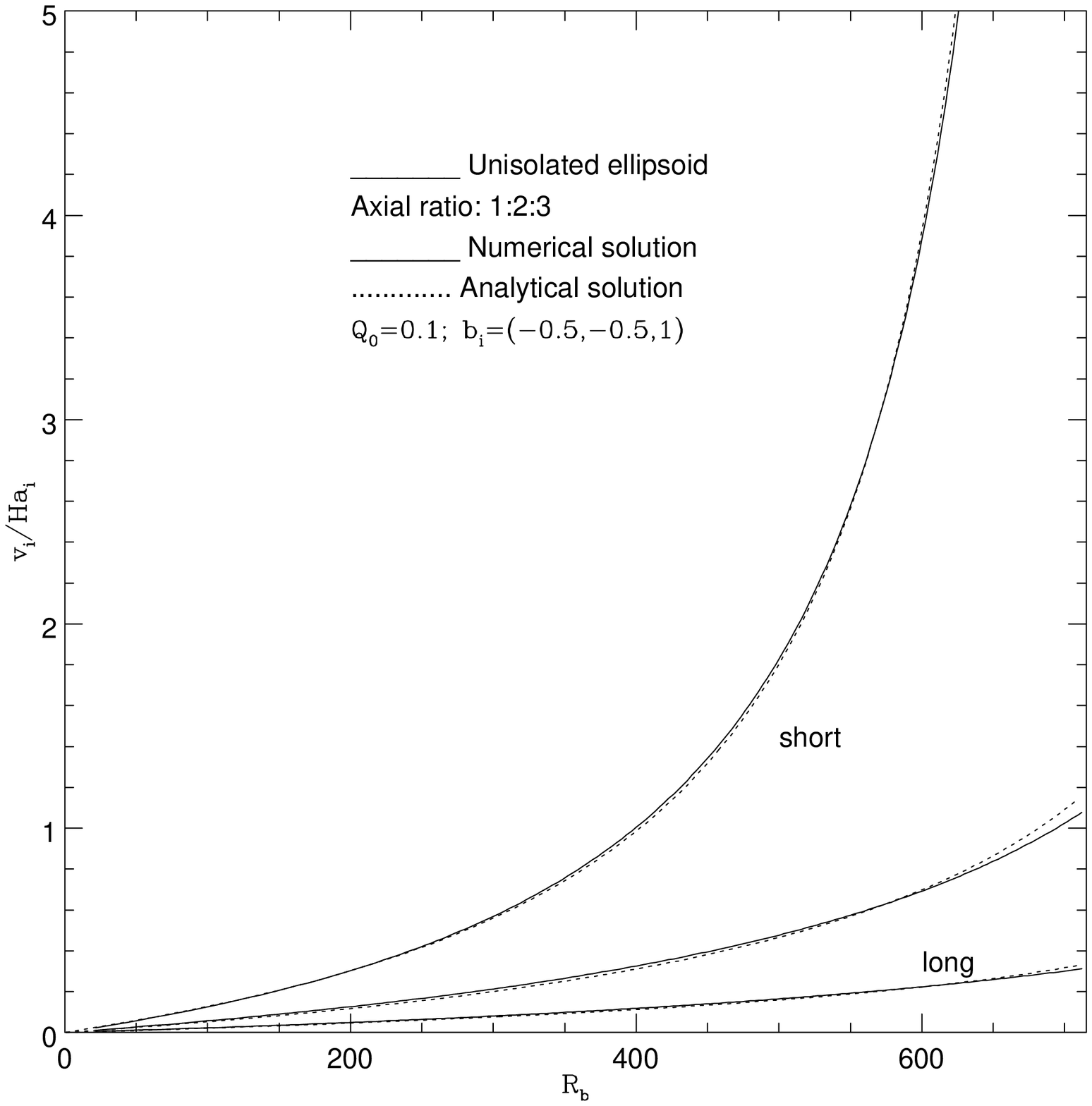,width=18cm}  }}
 {\bf Fig. 3c} Same as Fig. 3a but now the value of the axial ratio is $1:2:3$, while $Q_0=0.1$, and $b_{\rm
i}=(-0.5,-0.5,1)$. 
\end{figure}
\begin{figure}
\label{Fig. 3} \centerline{\hbox{
Fig. 3 (d)
\psfig{figure=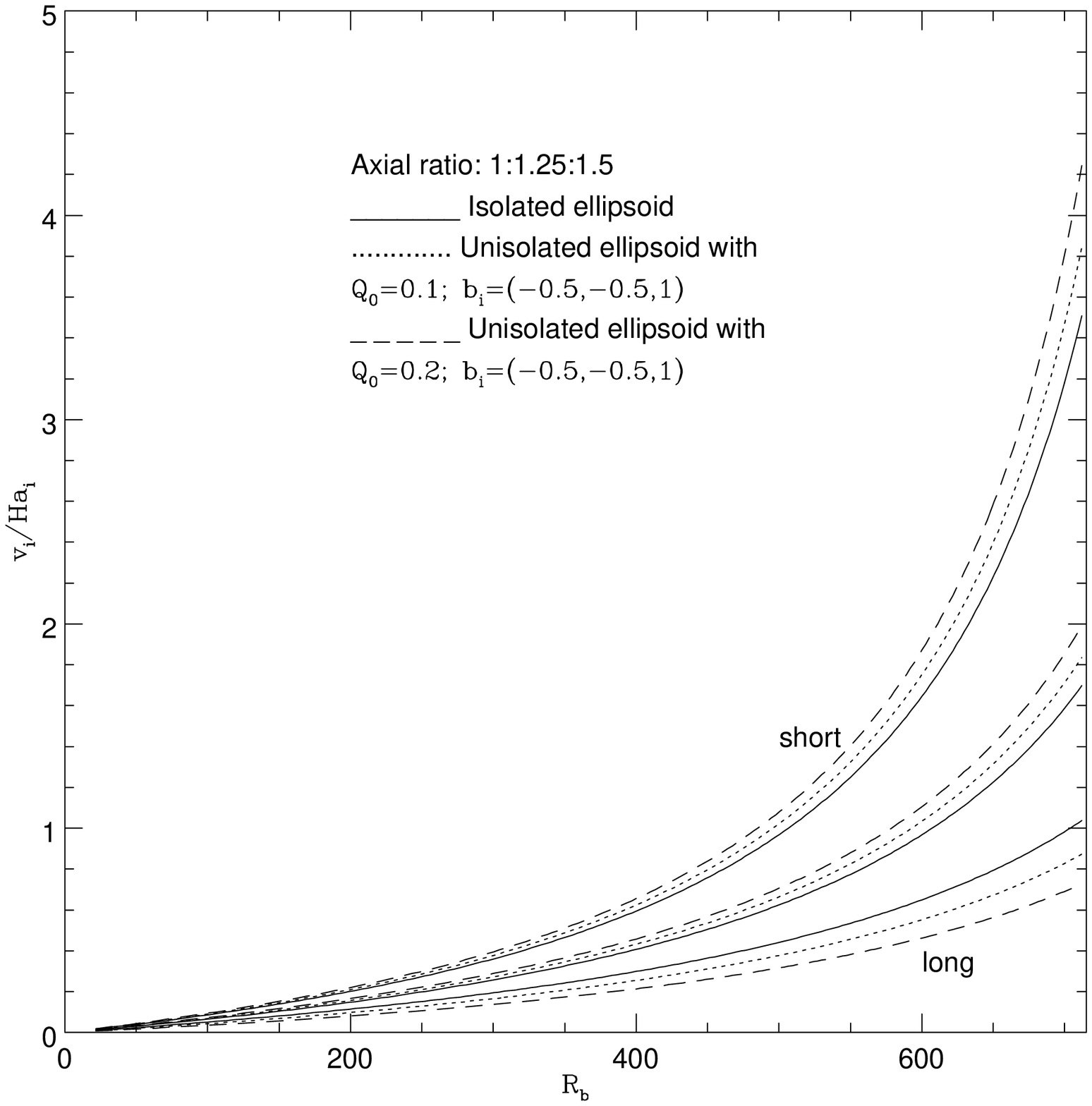,width=18cm}  }}
 {\bf Fig. 3d} Evolution of the axial velocity. The solid line represents the peculiar velocity
for an ellipsoid of axial ratio $1:1.25:1.5$ when no external
field is present, while the dotted and dashed lines represent the
case $Q_0=0.1$, $b_{\rm i}=(-0.5,-0.5,1)$, and $Q_0=0.2$, $b_{\rm i}=(-0.5,-0.5,1)$,
respectively.
\end{figure}
\begin{figure}
\label{Fig. 3} \centerline{\hbox{
Fig. 3 (e)
\psfig{figure=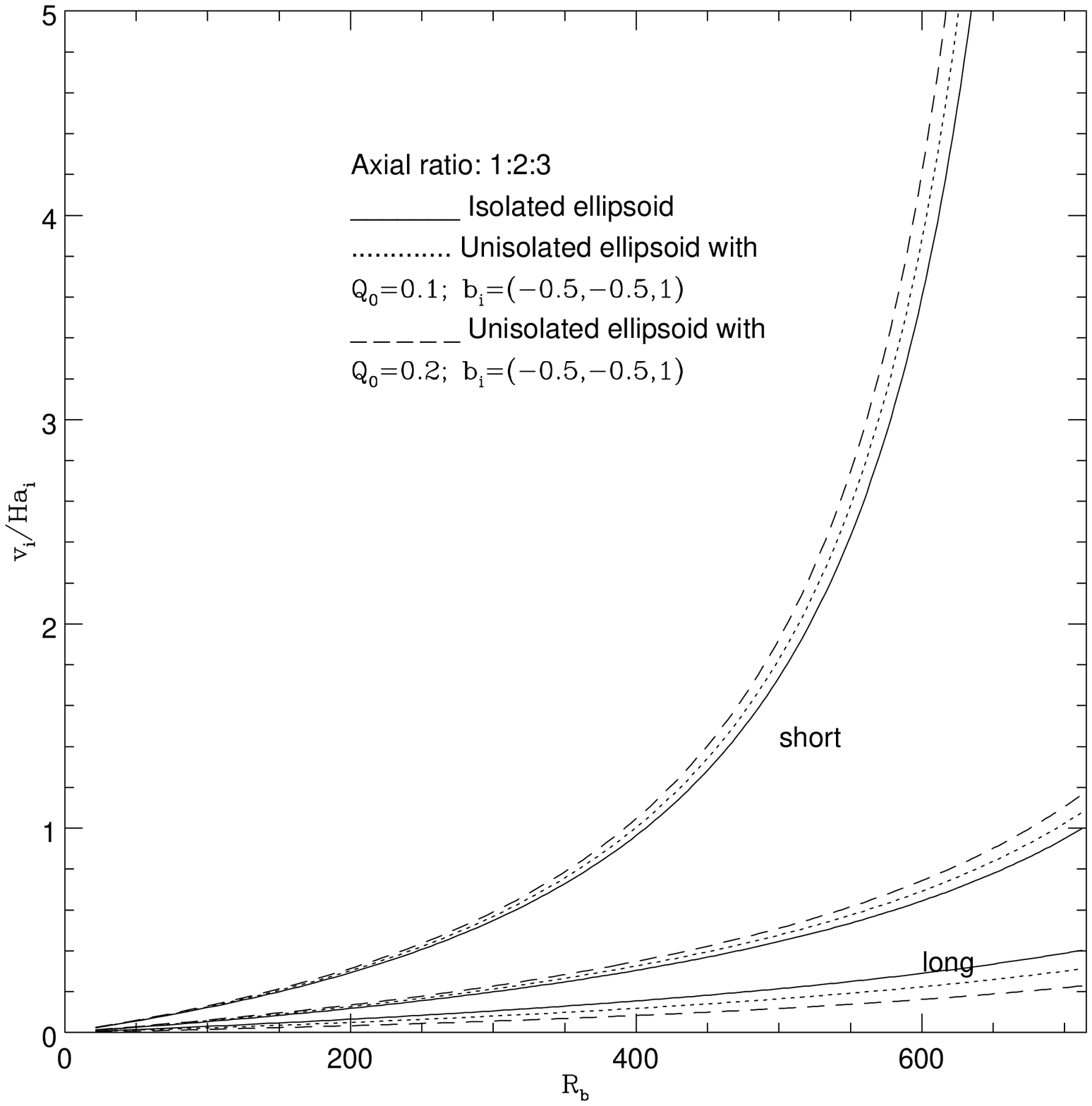,width=18cm}  }}
 {\bf Fig. 3e} Same as Fig. 3d, but now the axial ratio is $1:2:3$.
\end{figure}

Fig. 1a-1c, the ellipsoids have initial axial ratios of
$a_1:a_2:a_3=1:1.25:1.5$, $1:1.5:3$, $1:1.5:5$, respectively. The
solid lines represent the result of the numerical solutions and
the dotted lines the approximate analytical solution (Eq.
(\ref{eq:pred})). The top line represent the longest axis
evolution and the bottom one that of the shortest one. In this
section, I am principally interested in testing the goodness of
the approximation and less in describing the general properties of
the evolution of the perturbation, a point which was widely
discussed in WS and several other papers. However, two interesting
points, that emerge from the calculation are worth noting: \\
a) the shape of a perturbation is conserved until it acquires a
significant overdensity with respect to the background.\\
b) The collapse time of perturbations of fixed initial
overdensity, for a fixed background, decreases with increasing
initial asymmetry.
In other words the internal shear influences
the collapsing region geometry. More anisotropic initial
configurations are characterized by an acceleration of the
collapse along the shortest axis and a slowing down along the
longest one. As we shall see later, a similar effect is produced by
the external shear.

Coming back to the goodness of the approximation of Eq.
(\ref{eq:pred}) we have, for example, in the first case
($a_1:a_2:a_3=1:1.25:1.5$), that the error in $a_2(t_{\rm c})$ is
$\simeq 7\%$ while that in $a_3(t_{\rm c})$ is $\simeq 8\%$, where
$t_{\rm c}$ is the collapse time of the first axis. For a
configuration $a_1:a_2:a_3=1:1.5:3$, Fig. 1b shows that Eq.
(\ref{eq:pred}) gives a worse approximation going from less
asymmetric to more asymmetric configurations, especially in the
case of the shortest axis. This trend is confirmed by Fig. 1c,
representing the initial configuration $a_1:a_2:a_3=1:1.5:5$. The
same problem was encountered by Watanabe \& Inagaki (1991) in the
calculation of the axial peculiar velocity using WS approximate
solution.

It is possible to improve the approximation modifying 
Eq.(\ref{eq:pred}) slightly, and introducing some parameters whose
values can be obtained using the least-square method:
\begin{equation}
\frac{a_1(t)}{a_1(t_i)}=R_{\rm b}-\frac 32\alpha_1\left( R_{\rm
b}-R_{\rm e}\right)-d \times R_{\rm b}^{\left(\frac{2+3
c_1}{2}\right)} \left(1-\frac{3 \alpha_1}{2}\right)
\label{eq:pred1}
\end{equation}

\begin{equation}
\frac{a_{\rm 2}(t)}{a_{\rm 2}(t_i)}=R_{\rm b}-\frac 32 c_2 \alpha
_{\rm 2}\left( R_{\rm b}-R_{\rm e}\right) \label{eq:pred2}
\end{equation}

\begin{equation}
\frac{a_{\rm 3}(t)}{a_{\rm 3}(t_i)}=R_{\rm b}-\frac 32 c_3 \alpha
_{\rm 3}\left( R_{\rm b}-R_{\rm e}\right) \label{eq:pred3}
\end{equation}
For ellipsoids having an initial axial ratio $1:a_2:a_3$ with $a_1
\ge 1.25$ and $a_2 \ge 1.5$, we have that $d=7.22 \times 10^{-7}$
and $c_{\rm i}$ is given by:
\begin{equation}
c_1=a_{10}^{-0.115}a_{20}^{0.035}a_{30}^{0.08}, \hspace{0.5cm}
c_2=a_{10}^{0.07}a_{20}^{-0.06}a_{30}^{-0.01}, \hspace{0.5cm}
c_3=1.002 a_{10}^{0.1}a_{20}^{-0.035}a_{30}^{-0.065} \label{}
\end{equation}
where $a_{\rm i0}$ represents the initial value of the i-th axis.

Fig. 1d-1f shows that Eqs. (\ref{eq:pred1})-(\ref{eq:pred3})
give a better representation of numerical results, with respect to
Eq. (\ref{eq:pred}), for all of the three axes (the initial axial ratio
is the same as Figs. 1a-1c).

Similarly to the case of an {\it isolated} ellipsoid, it is
possible to obtain an analytical solution of Eq.
(\ref{eq:WSM}), describing the evolution of an {\it unisolated}
ellipsoid. In this case the solution can be written in the form:
\begin{equation}
\frac{a_1(t)}{a_1(t_i)}=R_{\rm b}-\frac 32 {\tilde \alpha_1}\left(
R_{\rm b}-R_{\rm e}\right)-d \times R_{\rm b}^{\left(\frac{2+3
c_1}{2}\right)} \left(1-\frac{3 {\tilde \alpha_1}}{2}\right)
\label{eq:predd1}
\end{equation}

\begin{equation}
\frac{a_{\rm 2}(t)}{a_{\rm 2}(t_i)}=R_{\rm b}-\frac 32 {\tilde
\alpha _{\rm 2}}\left( R_{\rm b}-R_{\rm e}\right)
\label{eq:predd2}
\end{equation}

\begin{equation}
\frac{a_{\rm 3}(t)}{a_{\rm 3}(t_i)}=R_{\rm b}-\frac 32 {\tilde
\alpha _{\rm 3}}\left( R_{\rm b}-R_{\rm e}\right)
\label{eq:predd3}
\end{equation}
For ellipsoids having initial axial ratio $1:a_2:a_3$, with $a_1
\ge 1.25$ and $a_2 \ge 1.5$, we now have that $c_1=1.23$,
$d=6\times10^{-7}$ and:
\begin{equation}
{\tilde \alpha _{\rm 1}}=\alpha_1+0.0672
\left(\frac{b_1}{b_2}\right)^{0.15} b^{0.6}_3 \label{}
\end{equation}
\begin{equation}
{\tilde \alpha _{\rm 2}}=a_{10}^{0.07}a_{20}^{-0.06}a_{30}^{-0.01}
\left[\alpha_2+0.031 \left(\frac{b_2}{b_1}\right)^{0.5} b^{0.95}_3
\right] \label{}
\end{equation}
\begin{equation}
{\tilde \alpha _{\rm 3}}=1.002
a_{10}^{0.1}a_{20}^{-0.035}a_{30}^{-0.065} \left(\alpha_3-0.063
a_{30}^{0.09} b^{0.95}_3 \right) \label{}
\end{equation}
where $b_{\rm i}$ was defined in Eq. (\ref{eq:bb}).

In the
case of prolate spheroids, with axial ratio $1:1:a_3$ and $1 \leq
a_3 \leq 5$, a better approximation to the $\alpha_{\rm i}$ is:
\begin{equation}
{\tilde \alpha _{\rm 1}}=\alpha_1+0.037 \left(
\frac{a_{30}}{a_{10}}
\right)^{0.35}
\left(\frac{b_1}{b_2}\right)^{0.15} b^{0.6}_3 \label{}
\end{equation}
\begin{equation}
{\tilde \alpha _{\rm 2}}=\left(
\frac{a_{10}}{a_{30}}
\right)^{0.01}
\left[\alpha_2+0.031
\left(
\frac{a_{10}}{a_{30}}
\right)
\left(\frac{b_2}{b_1}\right)^{0.5} b^{0.95}_3
\right] \label{}
\end{equation}
\begin{equation}
{\tilde \alpha _{\rm 3}}=1.002
\left(
\frac{a_{20}}{a_{30}}
\right)^{0.065} \left(\alpha_3-0.063
a_{30}^{0.09} b^{0.95}_3 \right) \label{}
\end{equation}

The comparison between the numerical results and Eqs.
(\ref{eq:predd1})-(\ref{eq:predd3}) are shown in Figs. 2a-2c (the
notation in Figs. 2a-2c is the same as fig. 1a-1f). In Figs.
2a-2c, I study the evolution of an ellipsoid having initial axes
in the ratio $1:1.25:1.5$ for different values of $Q_0$, and $b_{\rm
i}$ ($b_{\rm i}=(-0.5,-0.5,1)$, and $Q_0=0.1,0.2,0.4$).
These figures show that for a fixed initial configuration the
effect of the external tidal field causes a qualitatively
different evolution with respect to an {\it isolated} ellipsoid.
The ellipsoid expands less rapidly along the $x$ and $y$ axes, due
to the extra decelerating effect of the positive tidal components
($-b_{\rm i}=\left(0.5,0.5,-1 \right)$), and more rapidly along
the $z$ axis, since the tidal force is negative along this axis,
and this effect increases with increasing strength of the external
tidal field. Summarizing, the effect of a positive tidal force
along a given axis is to slow down the expansion of the
ellipsoid along this axis, and the effect of a negative tidal field
is to accelerate the expansion of the ellipsoid along that
axis. In other words,
shear
produce dilation in one direction and contraction in the other two
and vice versa (contraction in one direction and dilation in the
other two), according to its configuration. The result is in
agreement with van de Weygaert (1996) and Audit et al. (1997)
(note that Eq. (\ref{eq:WSM}) and Eq. (1) of van de
Weygaert (1996) differ for the sign of the tidal tensor
components, such that the evolution of the axes results inverted with
respect to our model). One may expect that the shear can change the
nature of the secondary infall, influencing the final mass
acquired by a peak and the density profiles of halos (Bond \&
Myers 1993; van de Weygaert \& Babul 1994).
This idea is confirmed by Figs. 2a-2c, which show that the shear
produces a decrease in the collapse time of perturbations of fixed
initial overdensity, for a fixed background, and that the effect
increases with increasing strength of the external field.
%
%
It is natural to expect that, for some shear configurations and 
strong enough external field, even the core region of a
perturbation can be induced to fragment, giving rise to different
halos (van de Weygaert \& Babul 1994), or, as shown by van de
Weygaert (1996) and Audit et al. (1997), 
the external tidal field can induce even a void to
collapse, if the external field is strong enough.

In Figs. 2d-2e, I plot the evolution of the ellipsoids for a fixed value
of $Q_0$ and $b_{\rm i}$ and for different values of the initial axial
ratio, in order to study the effect of the internal shear. In
Figs. 2d-2e, the axial ratio is $1:1.25:2$,
$1:1.25:5$, respectively, $Q_0=0.1$ and $b_{\rm i}=(-0.5,-0.5,1)$.
The plots show a situation similar to that seen in
the case of external shear: namely, the collapse time of the
perturbation decreases with increasing initial anisotropy.
While for {\it isolated} ellipsoids or for {\it unisolated} ones
with small values of $|\gamma|$, the axial ratio does not
appreciably change until the perturbation enters a strongly
nonlinear regime, (the self-gravity is dominant), for large values
of $|\gamma|$, the collapse is anisotropic even for not large
values of $\delta$. This means that the collapsing region geometry
is strongly influenced by the external shear. If the external
field is strong enough, the external shear is dominant in shaping
the region, with respect to internal shear,
in agreement with Eisenstein \& Loeb (1995),
Watanabe(1993).

As discussed in Sect.~ (2), in this paper, differently from Watanabe (1993), I assume
that protostructures have an initial asphericity. 
By means of this assumption,
I have the noteworthy advantage of being able to study the joint effect of ``internal and external shear". 
So it is interesting to analyse the contribution to the asphericity of real clusters of galaxies (e.g. LSC), 
coming from inner shear and external shear. 
In the case of an {\it isolated} ellipsoid the length of the uncollapsed axes at collapse can be 
obtained similarly to WS, by means of Eq. (\ref{eq:pred2})-(\ref{eq:pred3}):
\begin{equation}
\frac{a_{\rm 3}(t_{\rm c})}{a_{\rm 2}(t_{\rm c})}=\frac{a_{\rm 3}(t_{\rm i})}{a_{\rm 2}(t_{\rm i})}
\frac{R_{\rm b}-\frac 32 c_3 \alpha
_{\rm 3}\left( R_{\rm b}-R_{\rm e}\right)}{R_{\rm b}-\frac 32 c_2 \alpha
_{\rm 2}\left( R_{\rm b}-R_{\rm e}\right)}
\end{equation}
where $t_{\rm c}$ is obtained by $a_1(t_{\rm c})=0$. Notice that the approximation given in WS (Eq. 11 or Eq. 13) is not very accurate for extreme configurations.

In the case of an {\it unisolated} ellipsoid the length of the uncollapsed axes at collapse can be 
obtained by means of Eq. (\ref{eq:predd2})-(\ref{eq:predd3}):
\begin{equation}
\frac{a_{\rm 3}(t_{\rm c})}{a_{\rm 2}(t_{\rm c})}=\frac{a_{\rm 3}(t_{\rm i})}{a_{\rm 2}(t_{\rm i})}
\frac{R_{\rm b}-\frac 32 \tilde \alpha
_{\rm 3}\left( R_{\rm b}-R_{\rm e}\right)}{R_{\rm b}-\frac 32 \tilde \alpha
_{\rm 2}\left( R_{\rm b}-R_{\rm e}\right)}
\end{equation}

If we consider an initial {\it isolated} ellipsoid with $a_1:a_2:a_3= 1:1.25:1.5$, I get at collapse a ratio $\frac{a_{\rm 3}(t_{\rm c})}{a_{\rm 2}(t_{\rm c})} \simeq 1.9$. For the same initial axial ratio and $b=(-0.5,-0.5,1)$, $Q_0=0.1$, I get $\frac{a_{\rm 3}(t_{\rm c})}{a_{\rm 2}(t_{\rm c})} \simeq 2.5$ which produces an increase of $\simeq 34 \%$ on the previous ratio. Increasing the value of $Q_0$ to $0.2$ implies that $\frac{a_{\rm 3}(t_{\rm c})}{a_{\rm 2}(t_{\rm c})} \simeq 3.3$ and so an increase of $\simeq 76 \%$ on the previous ratio. If $Q_0=0.4$, I get $\frac{a_{\rm 3}(t_{\rm c})}{a_{\rm 2}(t_{\rm c})} \simeq 5.9$, indicating a dominant role of external shear in shaping structures for high $Q_0$. 
We can study the effect of varying the initial configuration of ellipsoids in presence of external shear.
In the case of an {\it isolated} ellipsoid with $a_1:a_2:a_3= 1:1.25:1.7$, I get  
$\frac{a_{\rm 3}(t_{\rm c})}{a_{\rm 2}(t_{\rm c})} \simeq 2.8$. If the ellipsoid is {\it unisolated} ($b=(-0.5,-0.5,1)$, $Q_0=0.1$), I get $\frac{a_{\rm 3}(t_{\rm c})}{a_{\rm 2}(t_{\rm c})} \simeq 3.7$. The isolated ellipsoid with $a_1:a_2:a_3= 1:1.25:2$ gives $\frac{a_{\rm 3}(t_{\rm c})}{a_{\rm 2}(t_{\rm c})} \simeq 4.6$ while the {\it unisolated} ($b=(-0.5,-0.5,1)$, $Q_0=0.1$) one gives $\frac{a_{\rm 3}(t_{\rm c})}{a_{\rm 2}(t_{\rm c})} \simeq 6$.
\begin{table}
\caption{Lengths of the uncollapsed axes at collapse for different values of external and internal shear}
\begin{tabular}{lll}
\hline 
$Q_{0}$ & $a_{1}:a_{2}:a_{3}$ & $\frac{a_{3(t_{c})}}{a_{2(tc)}}$ \\
\hline
0 & 1: 1.25: 1.5 & 1.9 \\
\hline
0.1 & 1: 1.25: 1.5 & 2.5 \\
\hline
0.2 & 1: 1.25: 1.5 & 3.3 \\
\hline
0.4 & 1: 1.25: 1.5 & 5.9 \\
\hline
\hline
$Q_{0}$ & $a_{1}:a_{2}:a_{3}$ & $\frac{a_{3(t_{c})}}{a_{2(tc)}}$ \\
\hline
0 & 1: 1.25: 1.7 & 2.8 \\
\hline
0.1 & 1: 1.25: 1.7 & 3.7 \\
\hline
0 & 1: 1.25: 2 & 4.6 \\
\hline
0.1 & 1: 1.25: 2 & 6\\
\hline
\end{tabular}
\end{table}

The previous one are only some examples (summarized in Table. 1) of how the model can be used to get information on the 
effect of external and internal shear on structures formation.

\section{Evolution of the axial peculiar velocity and of the density contrast}

The approximate solution obtained in the previous section
can be used to calculate the
evolution of the axial peculiar velocity. I use
Eqs.~(\ref{eq:predd1})-(\ref{eq:predd3}) to calculate the
peculiar velocity along the axis of the ellipsoid in units of the
Hubble expansion velocity, H:
\begin{equation}
\frac{v_{\rm i}}{H a_{\rm i}}=\left|\frac{\dot a_{\rm i}}{H a_{\rm
i}}-1\right| \label{eq:vel}
\end{equation}
The results of the calculation are plotted in Figs. 3a-3e. In
Figs. 3a-3c, I study how the internal shear influences the
evolution of the velocity. In all three figures, the solid line
represents, from up to bottom, the numerical result for the
peculiar velocity along the shortest, medium and longest axis. The
dotted line represents the velocity obtained from Eq.
(\ref{eq:vel}) using the approximation for the semiaxes (Eqs.
(\ref{eq:predd1})-(\ref{eq:predd3})). Going from Fig. 3a to Fig.
3c, the value of $Q_0$ is 0.1, $b_{\rm i}=(-0.5,-0.5,1)$ and the axial
ratio $1:1.25:1.5$, $1:1.25:5$, $1:2:3$, respectively.
%
%
For what concerns the goodness of the approximation, the plots
show that the approximate solution is in good agreement with the
numerical results. From the physical point of view, the plots
show that there is a considerable difference between the peculiar
velocities of the longest and shortest axes and in particular that
the shorter is the axis the larger is the velocity of collapse.
%

The effect of the external field is shown in Figs. 3d-3e. The
solid line in Fig. 3d represents the peculiar velocity for an ellipsoid
of axial ratio $1:1.25:1.5$ when no external field is present,
while the dotted and dashed lines represent the case $Q_0=0.1$, $b_{\rm
i}=(-0.5,-0.5,1)$, and $Q_0=0.2$, $b_{\rm i}=(-0.5,-0.5,1)$, respectively. The
situation is similar to that seen when I described the evolution
of the semiaxes for different values of the external field.
External shear produces different effects on the axes: the
evolution of the longest axis, which is characterized by the
smallest velocity, tend to be slowed down with increasing strength
of the external field (dotted and dashed lines). The effect of
external shear on the shortest and medium axes is opposite to that
on the longest one, namely shear produce an acceleration in their
evolution. Fig. 3e is the same as the previous one, but now the
axial ratio is $1:2:3$. Similarly to the previous plot, increasing
the strength of the external field produce an acceleration of
evolution in the shortest and medium axes and the opposite effect
on the longest one. Another feature shown by a comparison of Fig.
3d and Fig. 3e is that the evolution of the shortest axis is
accelerated with increasing asymmetry of the structure while the
opposite is true for the longest axis: as noted before, external
and internal shear have a qualitative similar effect on the
evolution of the shortest and longest axes. The difference between
the velocities along the longest and shortest axes is larger for
{\it unisolated} ellipsoids than for the {\it isolated} ones and
this difference increases with increasing strength of the external
field.

The evolution of the density contrast can be calculated using the
usual definition:
\begin{equation}
\delta =\frac{\rho _{\rm e}-\rho _{\rm b}}{\rho _{\rm
b}}=\frac{\rho _{\rm e0}}{\rho _{\rm
b0}}\frac{a_{10}}{a_1}\frac{a_{20}}{a_2}\frac{a_{30}}{a_3}\left(
\frac{R_{\rm b}}{R_{\rm b0}}\right) ^3-1 \label{eq:denss}
\end{equation}
\begin{figure}
\label{Fig. 4} \centerline{\hbox{
Fig. 4 (a)
\psfig{figure=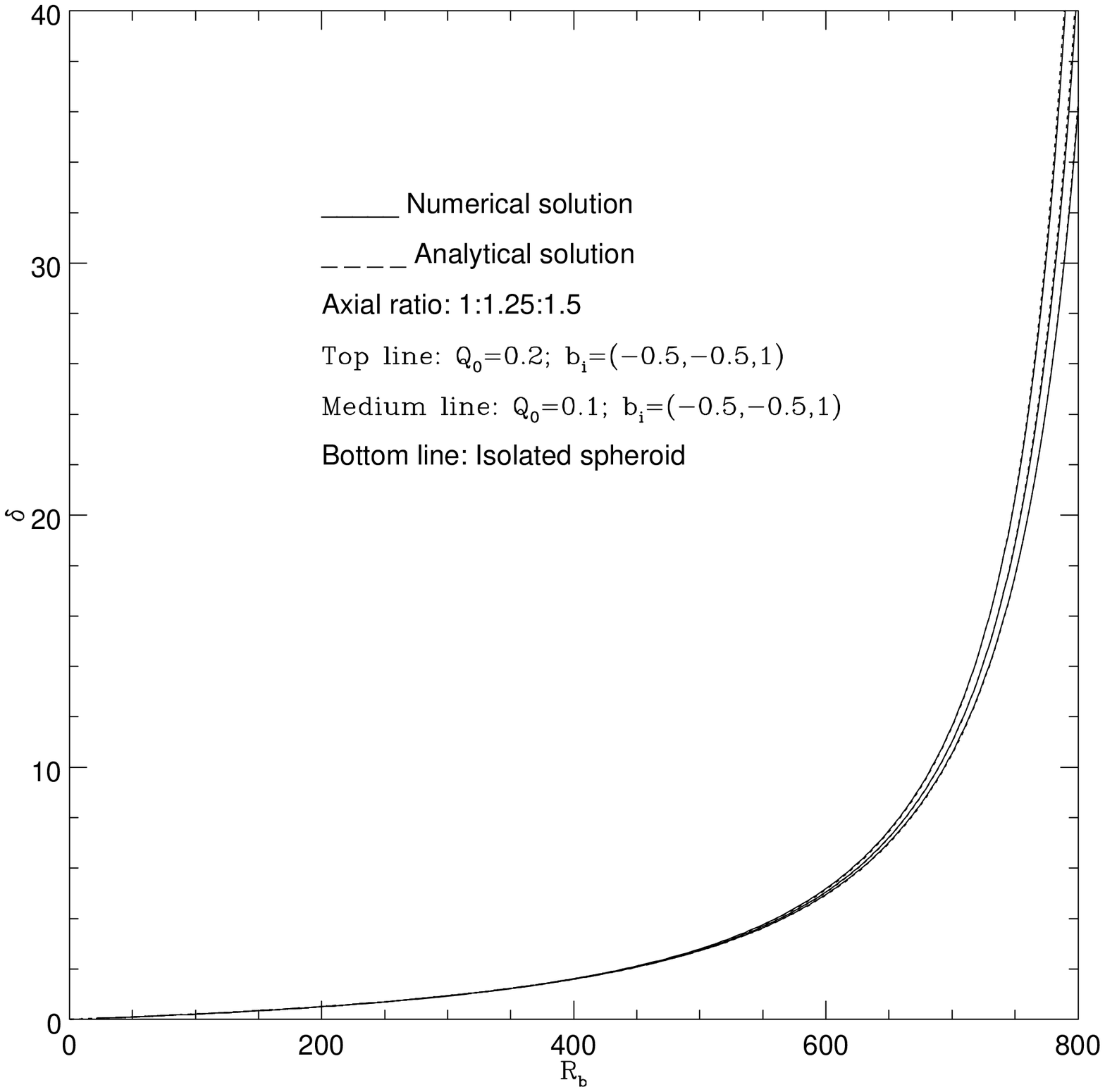,width=18cm}  }}
{\bf Fig. 4a} The evolution of the density contrast. The axial
ratio of the ellipsoid is $1:1.25:1.5$, the lines from bottom to
top represent the case of an {\it isolated} ellipsoid ($b_{\rm
i}=(0,0,0)$), $Q_0=0.1$, $b_{\rm i}=(-0.5,-0.5,1)$, and $Q_0=0.2$, $b_{\rm i}=(-0.5,-0.5,1)$,
respectively.
\end{figure}

\begin{figure}
\label{Fig. 4} \centerline{\hbox{
Fig. 4 (b)
\psfig{figure=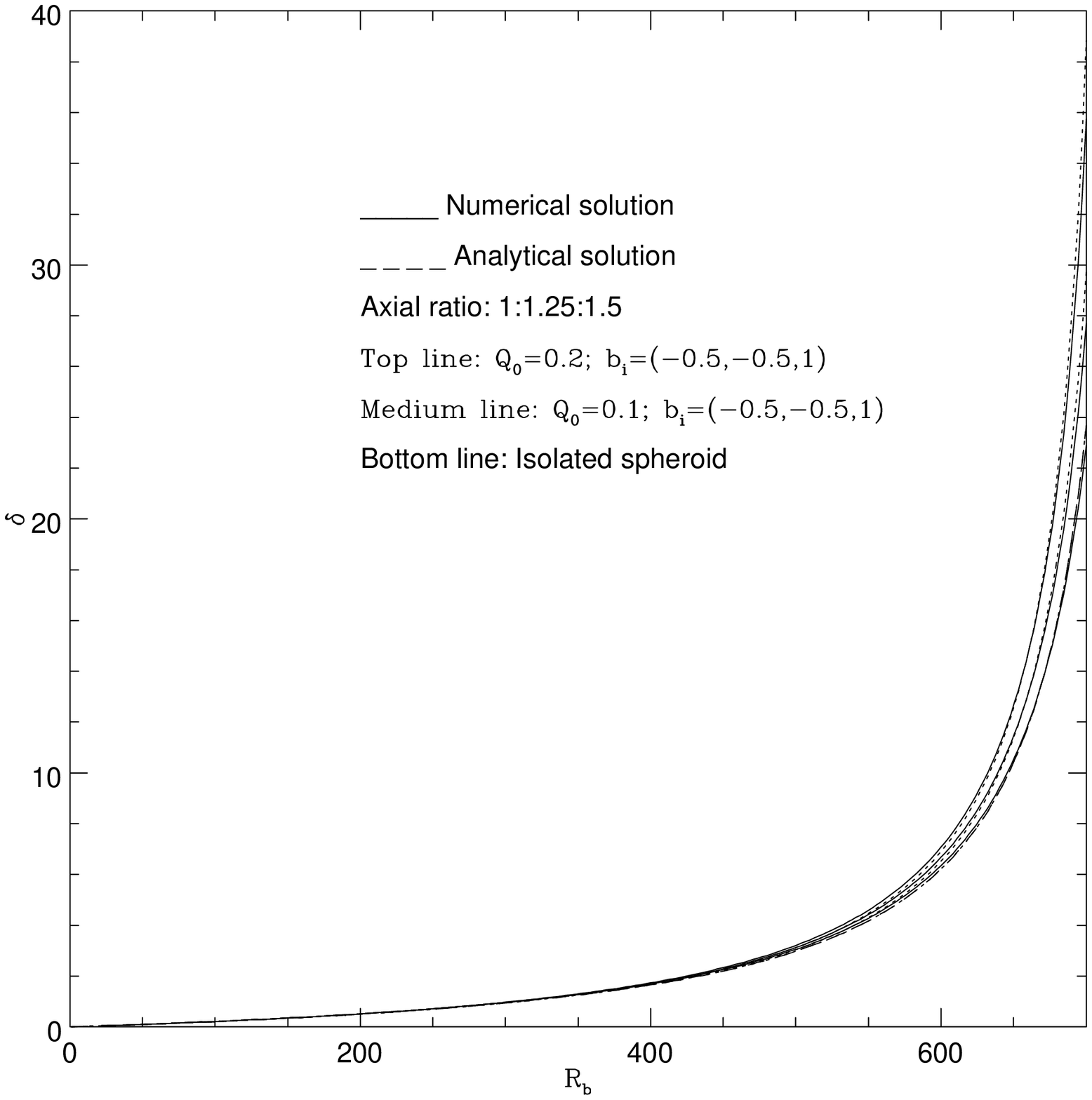,width=18cm}  }}
{\bf Fig. 4b} Same of the previous one but the configuration
is less spherical, with axial ratio $1:2:3$.
\end{figure}

\begin{figure}
\label{Fig. 5} \centerline{\hbox{
Fig. 5 
\psfig{figure=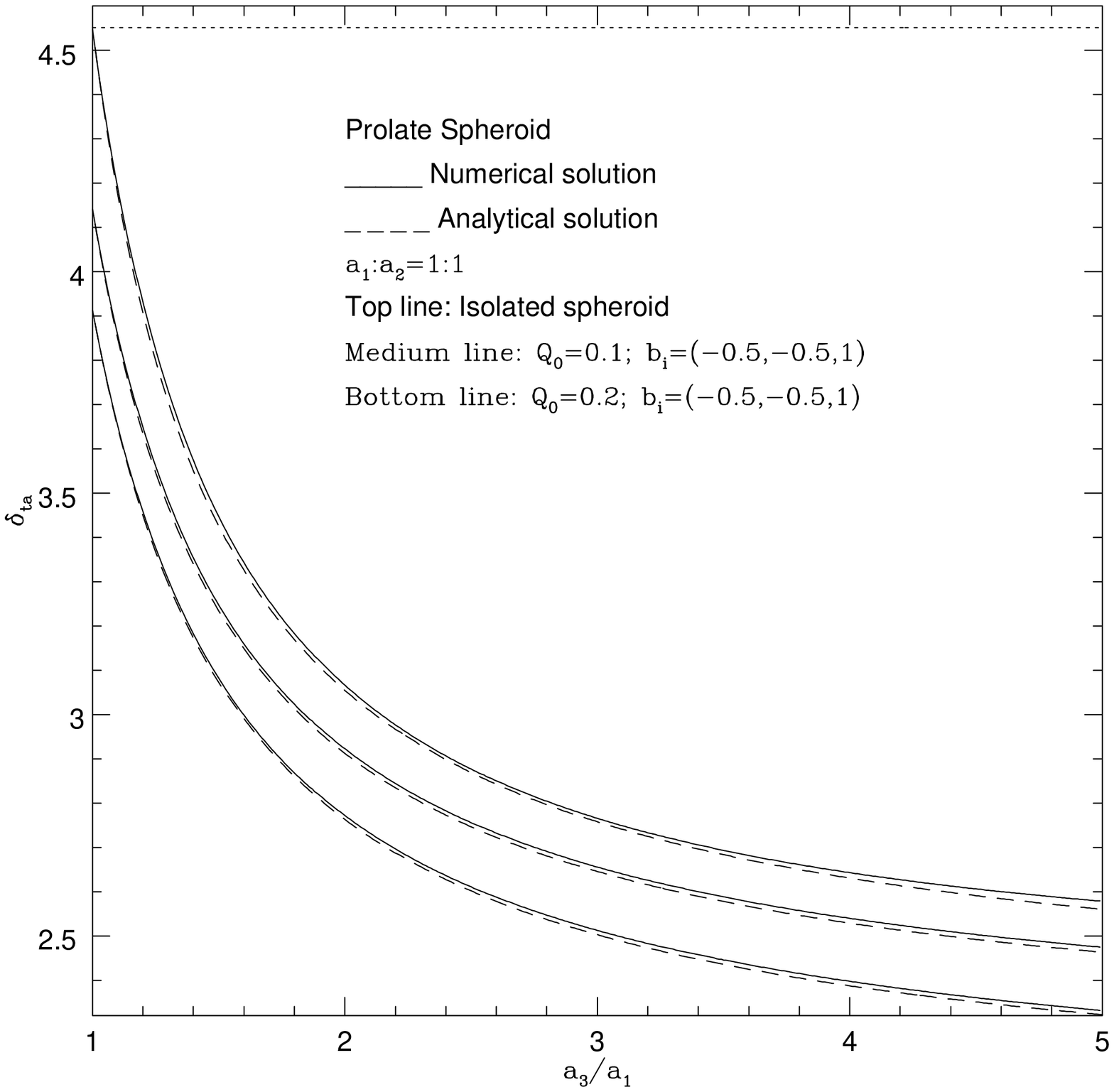,width=18cm}  }}
{\bf Fig. 5} The density contrast at turnaround for a prolate
spheroid for several values of the longest axis, $a_3$, (the other
two axes have fixed value $a_1:a_2=1:1$). The solid lines, from
top to bottom, represent numerical results for the density
contrast for an {\it isolated} spheroid ($b_{\rm i}=(0,0,0)$), and
for {\it unisolated} spheroids with $Q_0=0.1$, $b_{\rm i}=(-0.5,-0.5,1)$ and
$Q_0=0.2$, $(-0.5,-0.5,1)$, respectively. The dashed lines represent the
approximate solution (Eq. (37)). The upper
dotted line represents the value of the density contrast at
turnaround for a spherical perturbation.
\end{figure}

\begin{figure}
\label{Fig. 6} \centerline{\hbox{
Fig6. 
\psfig{figure=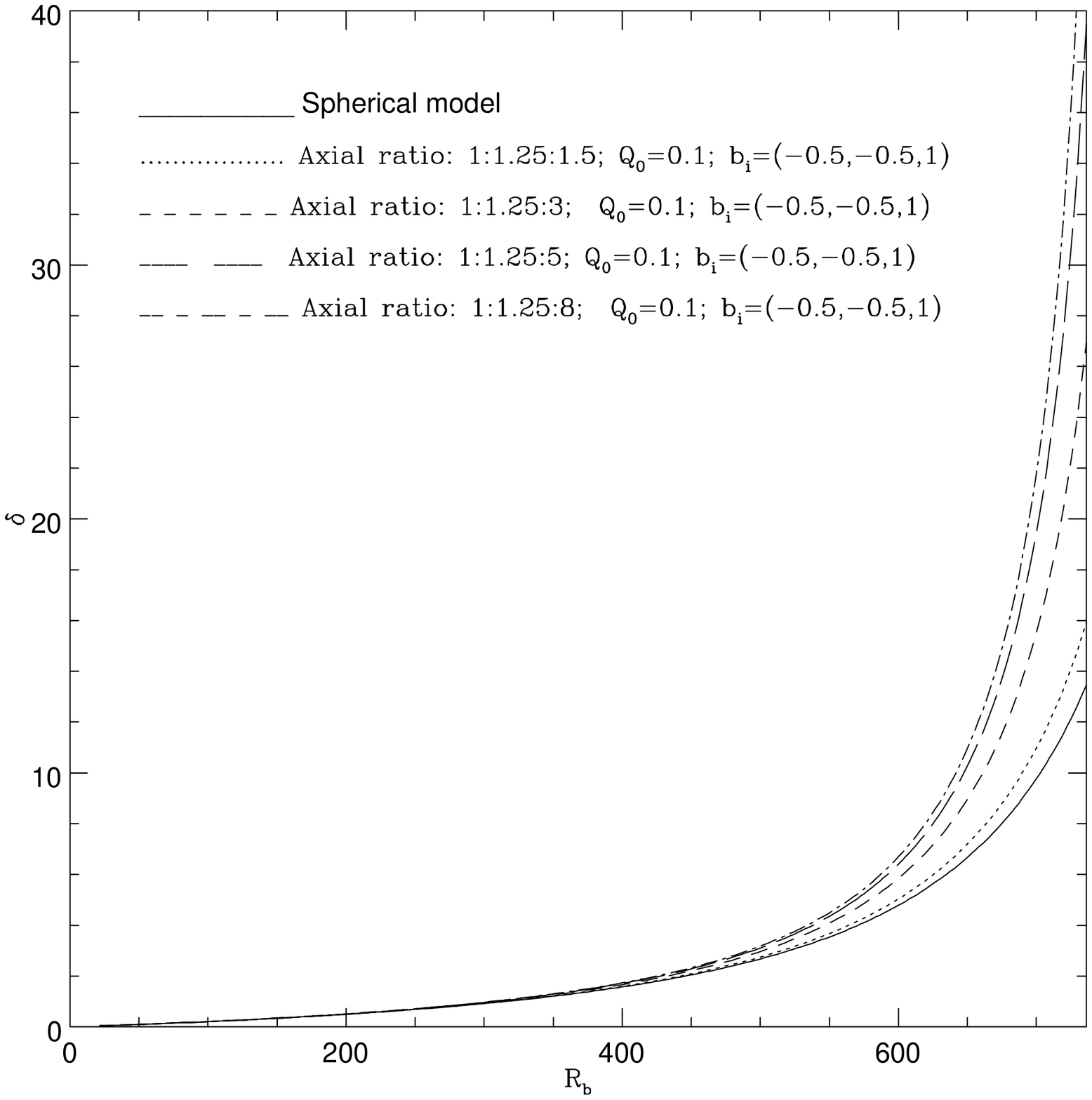,width=18cm}  }}
{\bf Fig. 6} Comparison between
density contrast predicted by spherical and ellipsoidal model. The
solid line represents the density contrast predicted by the
spherical model, while the dotted, short-dashed, long-dashed and
long-dashed--short-dashed lines represents the prediction of the
ellipsoid model with $Q_0=0.1$, $b_{\rm i}=(-0.5,-0.5,1)$ and initial
axial ratio $1:1.25:1.5$, $1:1.25:3$, $1:1.25:5$, and $1:1.25:8$
respectively.
\end{figure}

\begin{figure}
\label{Fig. 7} \centerline{\hbox{
Fig. 7
\psfig{figure=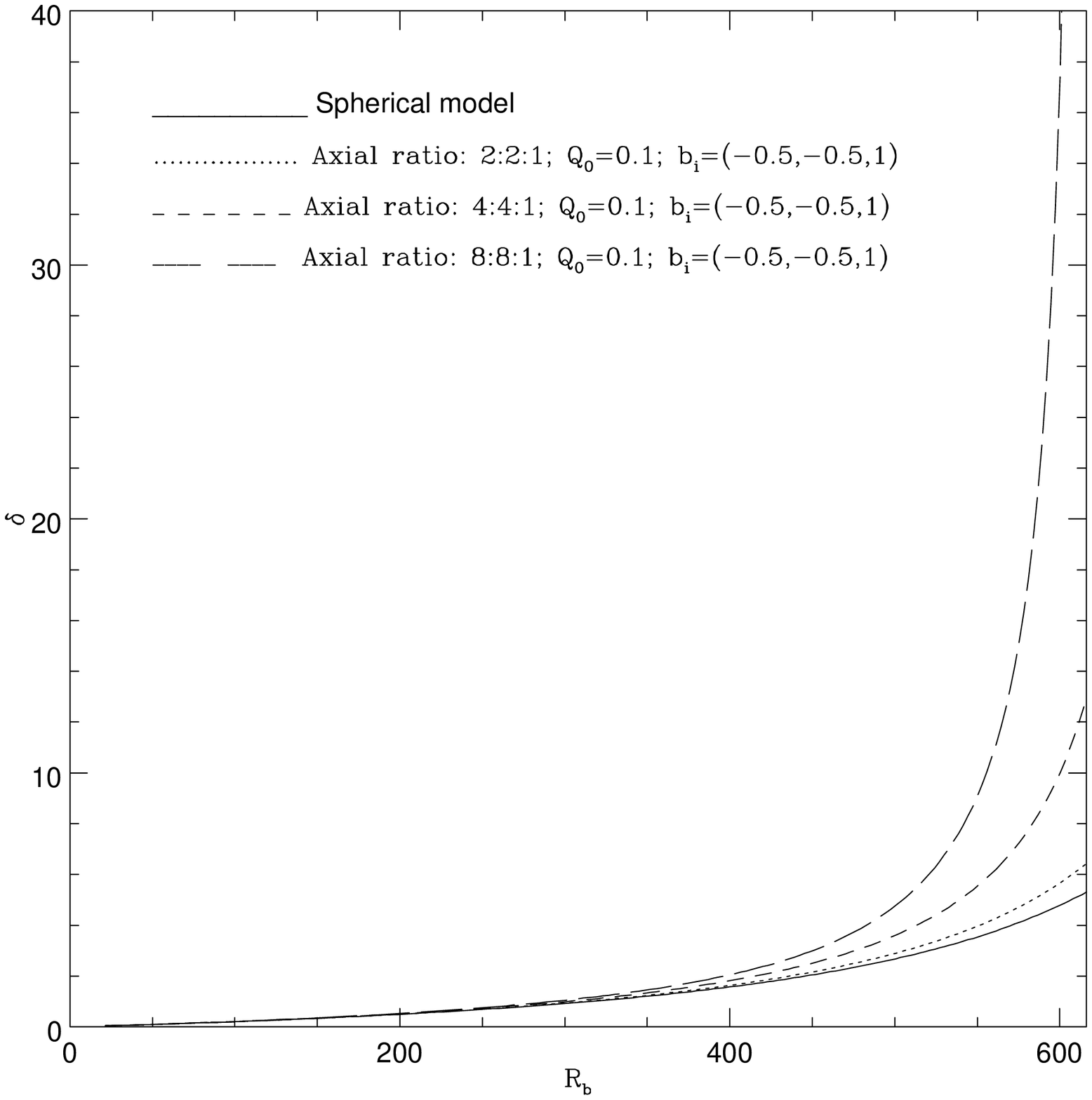,width=18cm}  }}
{\bf Fig. 7} Comparison
between density contrast predicted by spherical and oblate
spheroidal models. The solid line represents the density contrast
predicted by the spherical model, while the dotted, short-dashed,
and long-dashed lines represents the prediction of the model with
$Q_0=0.1$, $b_{\rm i}=(-0.5,-0.5,1)$ and initial axial ratio $2:2:1$,
$4:4:1$, $8:8:1$, respectively.
\end{figure}

\begin{figure}
\label{Fig. 8} \centerline{\hbox{
Fig. 8
\psfig{figure=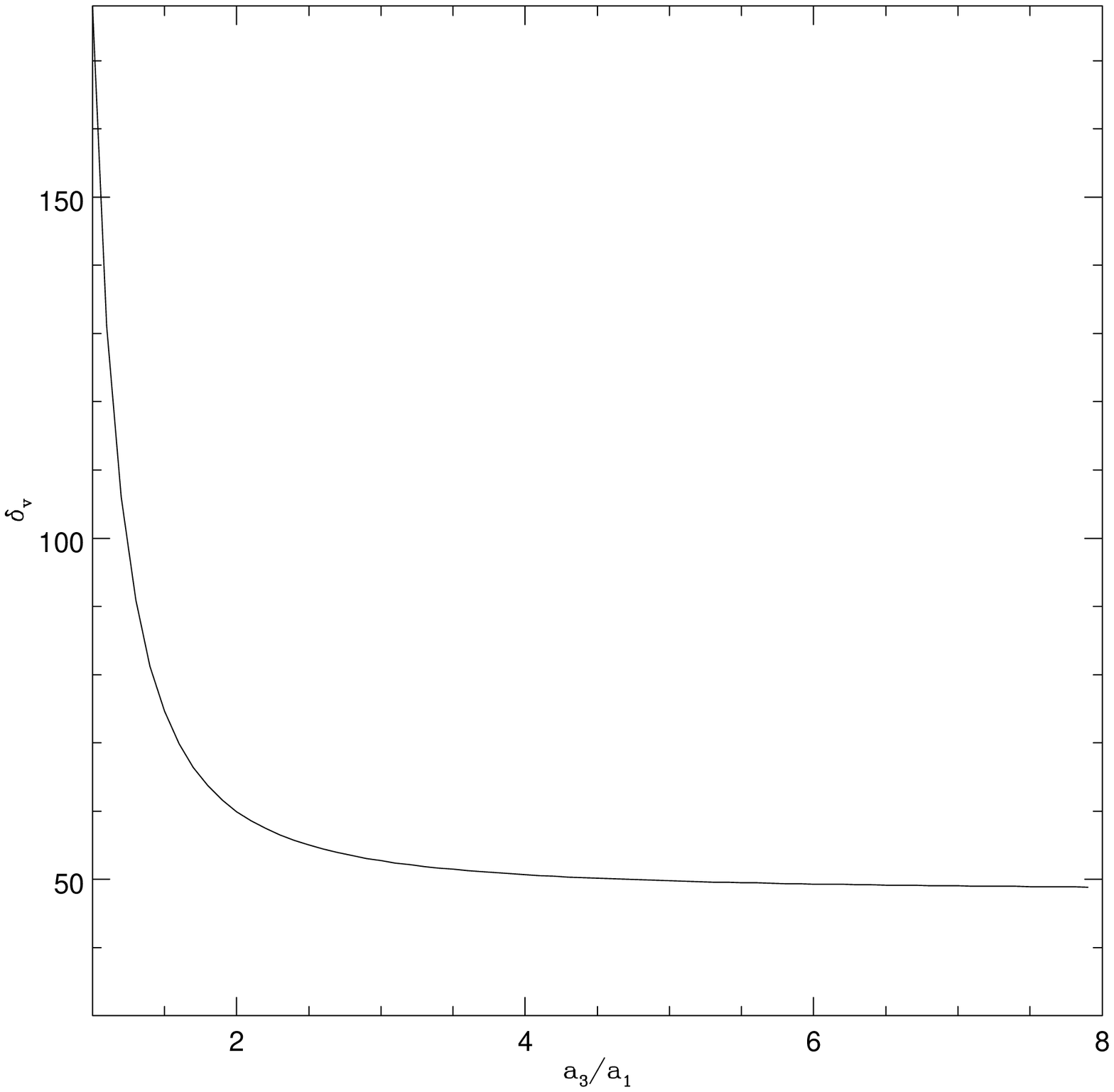,width=18cm}  }}
{\bf Fig. 8} Density contrast at virialization. The solid line
refers to an ${\it isolated}$ prolate spheroid.
\end{figure}

\begin{figure}
\label{Fig. 9} \centerline{\hbox{
Fig. 9
\psfig{figure=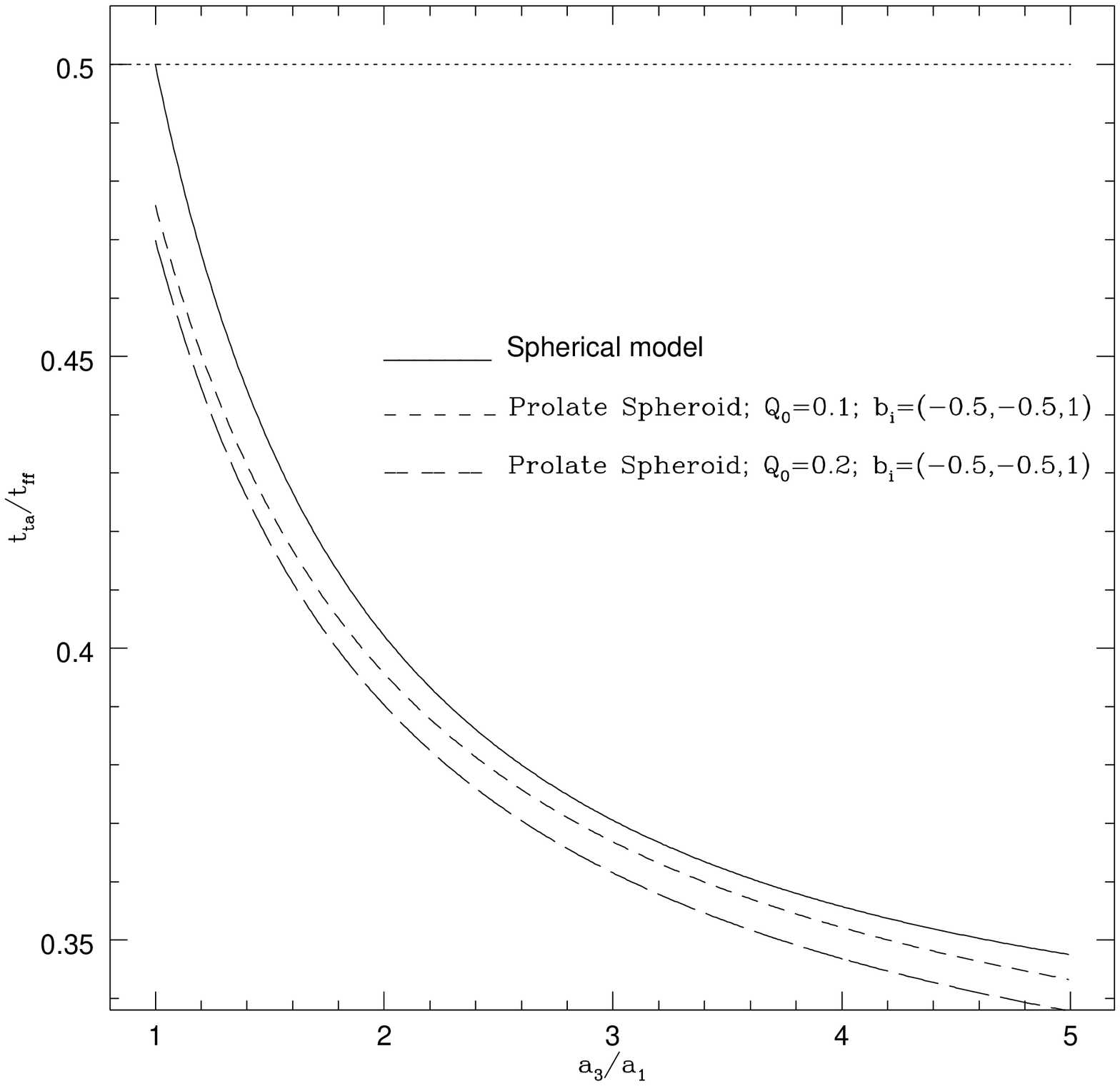,width=18cm}  }}
{\bf Fig. 9} Turnaround epoch for a prolate spheroid. The solid,
short-dashed and long-dashed lines, represent respectively the
time of turnaround for an {\it isolated} spheroid and for
spheroids having $Q_0=0.1$, $b_{\rm i}=(-0.5,-0.5,1)$ and $Q_0=0.2$, $(-0.5,-0.5,1)$. The
upper dotted line represents the value of the density contrast at
turnaround for a spherical perturbation.
\end{figure}

\begin{figure}
\label{Fig. 10} \centerline{\hbox{
Fig. 10
\psfig{figure=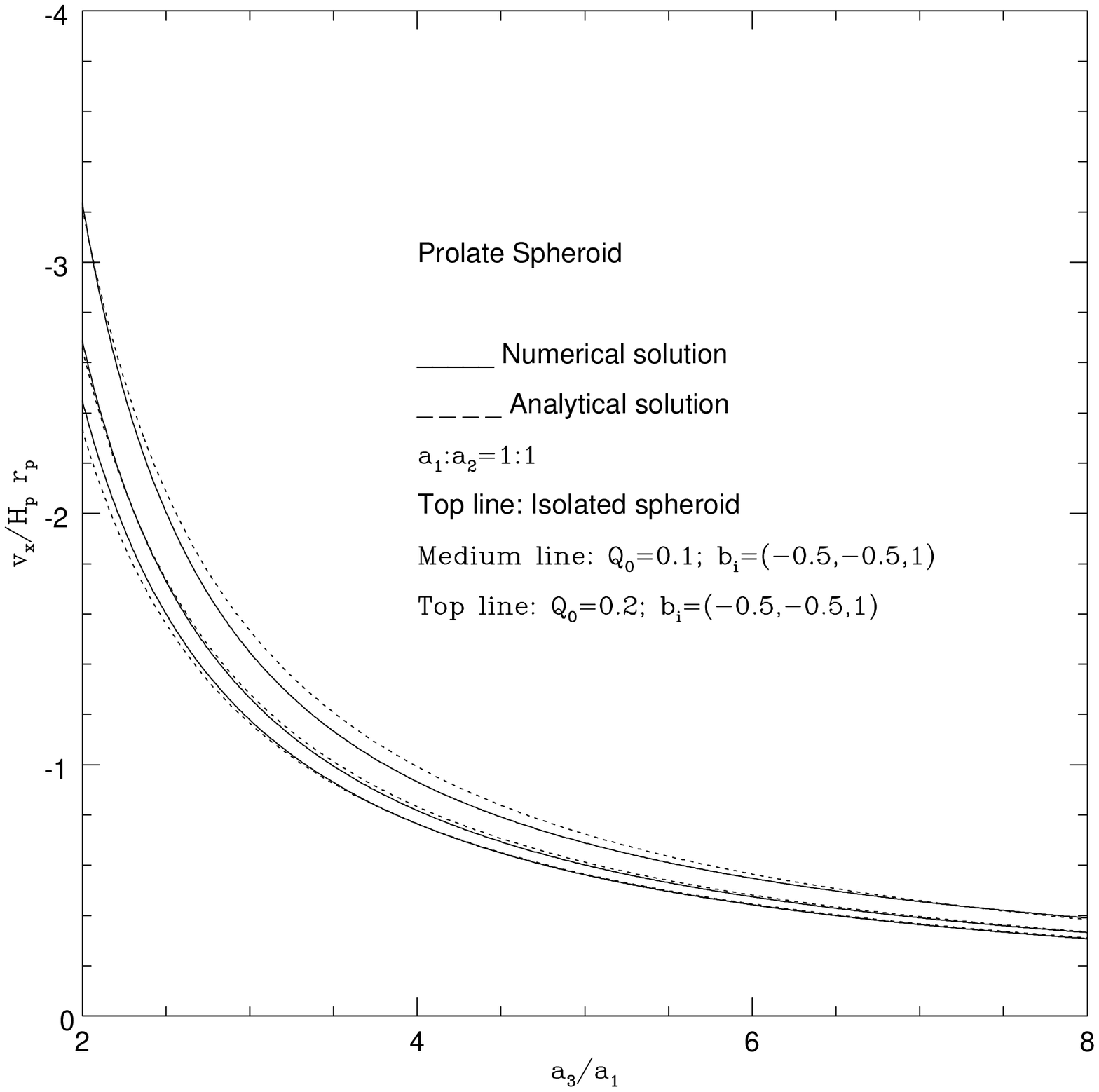,width=18cm}  }}
{\bf Fig. 10} Axial velocity at collapse as function of the ratio of
the initial value of the axes, $a_3/a_1$. The solid lines
represent the numerical results of the collapse velocity for a
prolate spheroid, the dotted lines the result obtained from the
approximate solution. The top curve represents the velocity for an
{\it isolated} spheroid, the medium one the same quantity for a
spheroid having $Q_0=0.1$, $b_{\rm i}=(-0.5,-0.5,1)$, and the last
bottom curve the velocity for a spheroid having $Q_0=0.2$, and $b_{\rm
i}=(-0.5,-0.5,1)$.
\end{figure}

As for the peculiar velocity, I calculated the evolution of the
density contrast of the ellipsoid both numerically and introducing
the approximate analytical solution (Eqs.
(\ref{eq:predd1})-(\ref{eq:predd3})) in Eq.
((\ref{eq:denss})). The result is plotted in Figs. 4a-4b. In both
figures the solid lines represent the numerical results and the
dotted lines the approximation, which is in good agreement with
the numerical results. In Fig. 4a, the axial ratio of the
ellipsoid is $1:1.25:1.5$, the lines from bottom to top represent
the case of an {\it isolated} ellipsoid ($b_{\rm i}=(0,0,0)$),
$Q_0=0.1$, $b_{\rm i}=(-0.5,-0.5,1)$, and $Q_0=0.2$, $b_{\rm i}=(-0.5,-0.5,1)$, respectively.
The plot shows that increasing the strength of the external field
the value of the density contrast increases. Fig. 4b is the same
of the previous one but the configuration is less spherical, with
axial ratio $1:2:3$. This last figure shows that more asymmetrical
initial configurations tend to have, at a given
time, larger values of $\delta$, which means that the collapse
time decreases with increasing initial asymmetry. Then internal
and external shear produce a more rapid evolution of the density
contrast.

\section{Overdensity at turnaround and velocity at collapse}

As reviewed in the introduction, in literature there is not full
agreement on the effect of shear on the collapse of density
perturbations: while according to Bertschinger \& Jain's (1994)
collapse theorem the spherical perturbations are the slowest in
collapsing, several other studies conclude in the opposite sense
(Peebles \& Groth 1976; Davis \& Peebles 1977; BS; Szalay \& Silk
1983; Villumsen \& Davis 1986; Peebles 1990; Bond \& Myers 1993
a,b; Lokas et al. 1996; Audit et al. 1997).\\
%
%
In other words,
the results concerning the effect of shear on collapse,
are of two opposite kind:\\
1) Shear slows down the collapse (Peebles \& Groth 1976; Davis \&
Peebles 1977; BS; Szalay \& Silk 1983; Villumsen \& Davis 1986;
Peebles 1990; Bond \& Myers 1993 a,b; Lokas et al. 1996; Audit et
al. 1997). This result is obtained, for example, if one uses
homogeneous ellipsoids to model an {\it extended} mass
distribution and not vanishing mass elements and collapse is
followed even after the collapse of the first axis: the collapse
of one axis is 'frozen' when it becomes smaller than a certain
value, and the collapse of the other axes is followed till the
collapse of the third axis, which defines the collapse time. This
is done to simulate the virialization process. In fact, as
remarked by Eisenstein and Loeb (1995), after the short axis
collapses, it makes a small contribution to the quadrupole moment
of the ellipsoid (the reason is due to the fact that the
quadrupoles are proportional to the difference between the squares
of the lengths of the axes) and then, in order to take 
into account the acquisition of angular momentum rightly, it is necessary
to follow the collapse after the first axis collapse.\\
2) Shear accelerates the collapse (e.g., Hoffmann 1986a, 1989;
Evrard \& Crone (1992), Bertschinger \& Jain (1994)). This result
is obtained in papers dealing with the evolution of individual
mass elements for which the collapse correspond to the collapse
of the first axis.

%
In the following of this section, I shall show that,
even if the effects of substructure are neglected, by
approximating the structure formation by means of an homogeneous
ellipsoid, and even if one assumes that the collapse is stopped
when the first axis collapses, the shear slows down the rate of
growth of the density contrast by lowering the peculiar velocity
(WS; BS; Szalay \& Silk 1983).

To this aim, in the following, I re-derive the fundamental
equations in BS for an homogeneous ellipsoid model taking also
into account the external field (in BS, only the
collapse of an {\it isolated} ellipsoid model was studied). These equations
shall be used to study the velocity at collapse and the
overdensity at turnaround.

The evolution of the ellipsoid can be obtained using Eqs.
(\ref{eq:predd1}-\ref{eq:predd3}):
\begin{equation}
x(t)=R_{\rm b}-\frac 32 {\tilde \alpha_1}\left( R_{\rm b}-R_{\rm
e}\right)-d \times R_{\rm b}^{\left(\frac{2+3 c_1}{2}\right)}
\left(1-\frac{3 {\tilde \alpha_1}}{2}\right) \label{eq:preddd1}
\end{equation}

\begin{equation}
y(t)=R_{\rm b}-\frac 32 {\tilde \alpha _{\rm 2}}\left( R_{\rm
b}-R_{\rm e}\right) \label{eq:preddd2}
\end{equation}

\begin{equation}
z(t)=R_{\rm b}-\frac 32 {\tilde \alpha _{\rm 3}}\left( R_{\rm
b}-R_{\rm e}\right) \label{eq:preddd3}
\end{equation}

Using BS notation, I indicate with $x(t)=x_{\rm o} X(t)$,
$y(t)=y_{\rm o} Y(t)$ and $z(t)=z_{\rm o} Z(t)$, the principal
axes ($x_{\rm o}$, $y_{\rm o}$ and $z_{\rm o}$ are the initial
values of the axes). The overdensity of the ellipsoid is the same
used till now, the initial conditions are $X=Y=Z=R_{\rm b}=R_{\rm
e}=1$ at $t=t_0$ and as before the initial velocity is equal to
the Hubble velocity at $t_0$ (representing the initial time). The
parametric equations satisfied by $R_{\rm e } (t)$ are:
\begin{equation}
R_{\rm e }=\frac{1}{2 \delta} \left(1-\cos(\vartheta)\right),
\hspace{0.5cm} \frac{t}{t_0}=\frac{3}{4
\delta^{3/2}}\left(\vartheta-\sin(\vartheta)\right) \label{eq:tim}
\end{equation}
%
while, since our background is an EdS universe, $R_{\rm b}(t)
\propto t^{2/3}$.

Following BS, it is easy to find that the density contrast
is given by:
\begin{eqnarray}
\delta_{\rm i} &=&\frac{R_{\rm b}^3}{XYZ}-1= f_1^2\left( \vartheta
\right) \left[ \left( 1-\frac{3{\tilde \alpha} _1}2\right)
f_1^{\frac 23}\left( \vartheta \right) +\frac 34{\tilde \alpha}
_1f_2\left( \vartheta \right) -d\left( 1-\frac{3{\tilde \alpha}
_1}2\right) \frac{f_1^{\frac{2f}3}\left( \vartheta
\right) }{\delta ^{f-1}}\right]^{-1} \times \nonumber \\
& & \left[ \left( 1-\frac{3{\tilde \alpha} _2}2\right) f_1^{\frac
23}\left( \vartheta \right) +\frac 34{\tilde \alpha} _2f_2\left(
\vartheta \right) \right]^{-1} \left[ \left( 1-\frac{3{\tilde
\alpha} _3}2\right) f_1^{\frac 23}\left( \vartheta \right) +\frac
34{\tilde \alpha} _3f_2\left( \vartheta \right) \right]^{-1} -1
\label{eq:denstu}
\end{eqnarray}
where $f=\frac{2+3c_1}{2}$, $f_1(\vartheta)=\frac{3}{4}
\left(\vartheta-\sin(\vartheta)\right)$ and
$f_2(\vartheta)=1-\cos(\vartheta)$. The density contrast at
turn-around is obtained by calculating $\delta_{\rm i}(\vartheta_{\rm
ta})$, where the parameter $\vartheta$ at turnaround epoch is
given solving the equation:
\begin{equation}
\frac 2{3 {\tilde\alpha_1} }=\frac{dfR^{f-1}_{\rm b}+\frac{\sin
(\vartheta_{\rm ta} )}{f_2(\vartheta_{\rm ta} )}f_1^{\frac
13}-1}{dfR^{f-1}_{\rm b }-1} \label{eq:time}
\end{equation}
Eqs. (\ref{eq:denstu}), (\ref{eq:time}) reduces to BS
Eqs. (76) and (72) for $d=0$, ${\tilde\alpha_1}=\alpha_1$,
${\tilde\alpha_2}=\alpha_2$, ${\tilde\alpha_3}=\alpha_3$.
\footnote{In reality some equations in BS contains some
typographical misprints, for example this is the case of Eq.
(76)  and Eq. (80).} Eq. (\ref{eq:denstu}) yields the
familiar value $\delta=\left(3 \pi/4\right)^2$  in the spherical
case,
($d=0$,
${\tilde\alpha_1}=\alpha_1={\tilde\alpha_2}=\alpha_2={\tilde\alpha_3}=\alpha_3=2/3$).
In general, in order to obtain the density contrast at turnaround,
one has first to solve Eq. (\ref{eq:time}) for $\vartheta$
for an arbitrary axial ratio and substitute the value in
Eq. (\ref{eq:denstu}). The time of turn-around can be
calculated by means of:
\begin{equation}
t=\frac{3 t_0}{4
\delta^{3/2}}\left(\vartheta-\sin(\vartheta)\right)=\frac{t_{\rm
ff}}{2 \pi} \left(\vartheta-\sin(\vartheta)\right) \label{}
\end{equation}
where $t_{\rm ff}$ is the free-fall time:
\begin{equation}
t_{\rm ff}=\frac{3 \pi}{2 \delta^{3/2}} t_0
\label{}
\end{equation}

In Fig. 5, I plotted the density contrast at turn-around for a
prolate spheroid for several values of the longest axis, $a_3$,
(the other two axes have fixed value $a_1:a_2=1:1$) (note that
from now on, and in the figures, $a_{\rm i}$ represents the initial value
of the i-th axis). The solid
lines, from top to bottom, represent numerical results for the
density contrast for an {\it isolated} spheroid ($b_{\rm
i}=(0,0,0)$), and for {\it unisolated} spheroids with $Q_0=0.1$, $b_{\rm
i}=(-0.5,-0.5,1)$ and $Q_0=0.2$, $(-0.5,-0.5,1)$, respectively. The dashed lines
represent the approximate solution
(Eq. (\ref{eq:denstu})). The figure shows that the density
contrast at turn-around is reduced, relative to the value
$\delta=\left(3 \pi/4\right)^2$ typical of spherical collapse when
$a_3>1$. This reduction increases with increasing initial
asymmetry of the spheroid. Although not shown, the reduction is
larger for oblate spheroids: in this case, for extreme initial
flattening (8:8:1) the value of $\delta_{\rm ta}$ can be reduced
to values as small as $\simeq 1$. The effect of the external field
is that of reducing the value of $\delta_{\rm ta}$ even more. An
interesting question that can arise because of the previous result
is the following: if at turn-around the density contrast predicted
by the ellipsoid model can be a factor 3 smaller than that of the
spherical model, then one should expect that the difference
between the two model starts at $\delta <\left(3 \pi/4\right)^2$.
Does the prediction of these two models match in the linear theory? In
order to answer this question, I compared the value of $\delta$
given by the ellipsoid model to that of the spherical one. The
result is plotted in Fig. 6. The solid line represents the density
contrast predicted by the spherical model, while the dotted,
short-dashed, long-dashed and long-dashed--short-dashed lines
represents the prediction of the ellipsoid model with $Q_0=0.1$,
$b_{\rm i}=(-0.5,-0.5,1)$ and initial axial ratio $1:1.25:1.5$,
$1:1.25:3$, $1:1.25:5$, $1:1.25:8$, respectively. As shown, the
difference between the prediction of spherical and ellipsoid model
starts at $\delta>1$. In Fig. 7, the same calculation is repeated
for an oblate spheroid. Here, the solid line represents the
density contrast predicted by the spherical model, while the
dotted, short-dashed, long-dashed lines represents the prediction
of the model with $Q_0=0.1$, $b_{\rm i}=(-0.5,-0.5,1)$ and initial
axial ratio $2:2:1$, $4:4:1$, $8:8:1$, respectively. In this case
the two models start to give different predictions at $\delta
\simeq 0.6$. So in both cases, the two models are different only
when we are far away from the linear regime.

The density contrast at virialization is a bit more difficult to
calculate. To begin with, it is important to recall the difference
between virialization and collapse: this last term describe a
state of the system in which the density approaches infinity,
while virialization is characterized by $|U|=2 K$, where U and K
are, respectively, the potential and kinetic energy. Only in the
case of perfectly spherical infall, collapse and virialization are
synonymous (although in the case of a bound system one follows
rapidly the other). In this case, the infall cannot be halted and
it proceeds towards a singularity, with all mass of the system
collapsing to a single point which means that the density
contrast becomes infinite. This result is physically unacceptable
and to prevent the system to reach this state it is necessary to
introduce, by hand, the assumption that the collapse halts when
spherical shells reach a particular radius $r_{\rm f}=r_{\rm
v}=r_{\rm ta}/2$, (where $r_{\rm f}$, $r_{\rm v}$ and $r_{\rm ta}$
are the final radius, the virialization radius and the turnaround
radius, respectively). \footnote{This major drawback of the
spherical model is removed by taking account of the angular momentum
in the equation of motion (see Engineer et al. 2000).} Then, in
the spherical infall model, the density contrast of the virializing
structure is calculated assuming that, after the shell collapses,
the final radius is $r_{\rm f}=r_{\rm v}=r_{\rm ta}/2$: this leads
to the conclusion that $\delta_{\rm v} =178$. The density contrast
of a non-spherical virialized system can be calculated using the
same definition given for the spherical system, namely:
\begin{equation}
\delta_{\rm v}=\frac{R_{\rm b}^3(t_{\rm
c})}{X(t_{1/2})Y(t_{1/2})Z(t_{1/2})}-1 \label{}
\end{equation}
where $t_{1/2}$ is the time at which the shortest axis has a
length 1/2 of the value at maximum expansion,
$x(t_{1/2})=x_{ta}/2$ (see Peebles 1990).

The time at collapse of the shortest axis can be obtained by means
of the second of Eq. (\ref{eq:tim}) once the parameter
$\vartheta_{\rm c}$ at the collapse time, characterized by
$X(t_{\rm c})=0$, is known. After some calculations it is possible
to show that $\vartheta_{\rm c}$ can obtained solving the
following equation:
\begin{equation}
\frac 2{3{\tilde \alpha_1} }=\frac{\frac{1-\cos (\vartheta_{\rm c}
)}2f_1^{\frac{-2}3}\left( \vartheta_{\rm c} \right) +d\left(
\frac{f_1^{\frac 23}\left( \vartheta_{\rm c} \right) }\delta
\right) ^{f-1}-1}{d\left( \frac{f_1^{\frac 23}\left(
\vartheta_{\rm c} \right) }\delta \right) ^{f-1}-1}
\label{eq:tempp}
\end{equation}
which, for $d=0$ and ${\tilde \alpha_1}=\alpha_1$, reduces to
Eq. (78) of BS. The time $t_{1/2}$, or equivalently,
$\vartheta_{1/2}$, is obtained solving the equation:
\begin{equation}
x(t_{1/2})=x_{\rm ta}/2
\end{equation}
where $x(t)$ is given in Eq. (\ref{eq:preddd1}) and $x_{\rm
ta}$ is obtained using again Eq. (\ref{eq:preddd1}) and
calculating $x(\vartheta_{\rm ta})$.
%
%
The result is plotted in Fig. 8. The solid line shows that, with
increasing initial anisotropy, the density contrast decreases from
the value $\delta_{\rm v} =178$, for the spherical case, to a
value of $\delta_{\rm v}  \simeq 48$, for large initial
'flattening' (semiaxes ratio 1:1:8). The result plotted refers to
an ${\it isolated}$ prolate spheroid, while for the oblate case
and for the same axial ratio (8:8:1) results $\delta_{\rm v}
\simeq 11$. These results are in qualitative agreement with
Peebles (1990).

The results in Figs. 5-8 could seem in disagreement with Fig.
4a-4b, since in those figures the value of $\delta$ increases with
increasing initial asymmetry of the ellipsoid. The reason why
$\delta_{\rm ta}$ decreases with increasing initial asymmetry is
due to the fact that the turn-around $t_{\rm ta}$, and also the
collapse epoch, moves towards lower values of time, $t$, for
larger values of initial asymmetry of a given perturbation and
larger strength of the external field. This last effect is shown
in Fig. 9: the solid, short-dashed and long-dashed lines,
represent respectively the time of turn-around for an {\it
isolated} spheroid and for spheroids having $Q_0=0.1$, $b_{\rm
i}=(-0.5,-0.5,1)$ and $Q_0=0.2$, $(-0.5,-0.5,1)$. More asymmetric structures are
characterized by a smaller value of turn-around time, and the
external field contribute to this reduction of $t_{\rm ta}$. In
the case of oblate spheroids, the collapse time can be reduced to
values $t \simeq 0.3 t_{\rm ff}$ for initial axial ratio $8:8:1$.

Another interesting quantity that can be obtained is the collapse
velocity at the time of collapse. I shall calculate the collapse
velocity using the same steps followed by BS, with the difference
that the spheroids considered in the following are prolate. Their
evolution is obtained as before, putting Eqs.
(\ref{eq:preddd1}-\ref{eq:preddd3}) in the equation for the
collapse velocity down the shortest axis:
\begin{equation}
v_{\rm x}=x_0 \dot X \label{eq:vx}
\end{equation}
Calculating the derivative of Eq. (\ref{eq:preddd1}) and
inserting it in Eq. (\ref{eq:vx}), I find that:
\begin{equation}
v_{\rm xc}=-x_0 \frac 32{\tilde \alpha_1} \dot R_{\rm b}\left[
1-\frac 2{3{\tilde \alpha_1} }-\left( \frac 34\right) ^{\frac
13}\frac{\sin (\vartheta_{\rm c} )\left( \vartheta_{\rm c} -\sin
(\vartheta_{\rm c} )\right) }{1-\cos (\vartheta_{\rm c}
)}+dfR^{f-1}_{\rm b}\left( \frac 2{3{\tilde \alpha_1} }-1\right)
\right] \label{eq:velx}
\end{equation}
here the subscript ``c" means that the corresponding quantity is calculated
at the collapse epoch.
%
%
Eq. (\ref{eq:velx}) can be written in a different form using
the following equations:
\begin{equation}
Z_{\rm c}=R_{\rm bc}-\frac{3{\tilde \alpha_3} }2\left( R_{\rm
bc}-R_{\rm ec}\right) =R_{\rm bc}\left( 1-\frac {\tilde \alpha_3}
{\tilde \alpha_1} \right) +\frac{3{\tilde \alpha_3} }2d\left(
\frac 2{3{\tilde \alpha_1} }-1\right) R_{\rm b}^f \simeq R_{\rm
bc}\left( 1-\frac {\tilde \alpha_3} {\tilde \alpha_1} \right)
\label{}
\end{equation}
(see Eq. (79) of BS). Approximating Eq. (\ref{eq:tempp})
by means of Eq. (78) of BS:
\begin{equation}
1-\frac 2{3{\tilde \alpha_1} }=\frac{1-\cos (\vartheta_{\rm c}
)}2\left[ \frac 34\left( \vartheta_{\rm c} -\sin (\vartheta_{\rm
c} )\right) \right] ^{\frac{-2}3} \label{}
\end{equation}
and approximating the equation $x=0$, defining the collapse time,
with the Eq. (77) of BS:
\begin{equation}
\left(1-\cos(\vartheta_{\rm c})\right)/2 \delta=R_{\rm bc}
\left(1-\frac{2}{3 \alpha_1}\right) \label{}
\end{equation}
Finally I get:
\begin{eqnarray}
v_{\rm xc}&=&-\left[\frac{\dot R_{\rm bc}}{R_{\rm bc}}r_{\rm
c}\right]\frac{3{\tilde \alpha_1} -2}{{\tilde \alpha_1} -{\tilde
\alpha_3}} 
\{1-\frac 32\frac{\sin (\vartheta_{\rm c} )\left(
\vartheta_{\rm c} -\sin (\vartheta_{\rm c} )\right) }{\left(
1-\cos (\vartheta_{\rm c} )\right) ^2} +\nonumber \\
& & 2df\left[
\frac{1-\cos
(\vartheta_{\rm c} )}{2\delta \left( 1-\frac 2{3{\tilde \alpha_1}
}\right) }\right] ^{f-1}    
\left( \frac 2{3{\tilde \alpha_1}
}-1\right) \frac{\left[ \frac 34\left( \vartheta_{\rm c} -\sin
(\vartheta_{\rm c} )\right) \right] ^{\frac 23}}{1-\cos
(\vartheta_{\rm c} )} \} \frac{x_0}{z_0}{\tilde \alpha_1} \label{eq:vxx}
\end{eqnarray}
where $r_{\rm c}=z_0 Z_{\rm c}$ is the radius of the collapsed
object and $R_{\rm bc}$ represents $R_{\rm b}(t_{\rm c})$. Using
Eq. (\ref{eq:vxx}) instead of Eq. (\ref{eq:velx})
introduces errors of order $\simeq 10\%$. In the case of a prolate
spheroid, one has that:
\begin{eqnarray}
v_{\rm xc}&=&-\left[\frac{\dot R_{\rm bc}}{R_{\rm bc}}r_{\rm
c}\right]\frac{3{\tilde \alpha_1} -2}{{\tilde \alpha_1} -{\tilde
\alpha_3} } 
\{ 1-\frac 32\frac{\sin (\vartheta_{\rm c} )\left(
\vartheta_{\rm c} -\sin (\vartheta_{\rm c} )\right) }{\left(
1-\cos (\vartheta_{\rm c} )\right) ^2}+ \nonumber \\
& & 2df\left[ \frac{1-\cos
(\vartheta_{\rm c} )}{2\delta \left( 1-\frac 2{3{\tilde \alpha_1}
}\right) }\right] ^{f-1}\left( \frac 2{3{\tilde \alpha_1}
}-1\right) \frac{\left[ \frac 34\left( \vartheta_{\rm c} -\sin
(\vartheta_{\rm c} )\right) \right] ^{\frac 23}}{1-\cos
(\vartheta_{\rm c} )} \} \sqrt{1-e^2}{\tilde \alpha_1}
\label{}
\end{eqnarray}
where I have defined $1-e^2=\left(x_0/z_0\right)^2$. In the case
of an oblate spheroid:
\begin{eqnarray}
v_{\rm zc}&=&-\left[\frac{\dot R_{\rm bc}}{R_{\rm bc}}r_{\rm
c}\right]\frac{3{\tilde \alpha_3} -2}{{\tilde \alpha_3} -{\tilde
\alpha_1} }\{1-\frac 32\frac{\sin (\vartheta_{\rm c} )\left(
\vartheta_{\rm c} -\sin (\vartheta_{\rm c} )\right) }{\left(
1-\cos (\vartheta_{\rm c} )\right) ^2}+ \nonumber \\
& & 2df\left[ \frac{1-\cos
(\vartheta_{\rm c} )}{2\delta \left( 1-\frac 2{3{\tilde \alpha_3}
}\right) }\right] ^{f-1}\left( \frac 2{3{\tilde \alpha_3}
}-1\right) \frac{\left[ \frac 34\left( \vartheta_{\rm c} -\sin
(\vartheta_{\rm c} )\right) \right] ^{\frac 23}}{1-\cos
(\vartheta_{\rm c} )} \} \sqrt{1-e^2}{\tilde \alpha_3}
\label{}
\end{eqnarray}
having defined $1-e^2=\left(z_0/x_0\right)^2$ and $r_{\rm c}=x_0
X_{\rm c}$. If $d=0$,
${\tilde\alpha_1}=\alpha_1={\tilde\alpha_2}=\alpha_2={\tilde\alpha_3}=\alpha_3=2/3$,
the spheroid is isolated and Eq. (82)-(83) of BS is
recovered.

In Fig.~10, I plot $\frac{v_{{\rm x_c}}}{H_{\rm c} r_{\rm c}} $,
as a function of the ratio of the initial value of the axes,
$a_3/a_1$. The solid lines represent numerical results of the
collapse velocity for a prolate spheroid ($a_1=a_2<a_3$), the
dotted lines represent the result obtained from the approximate solution.
The top curve represents the velocity for an {\it isolated}
spheroid, the medium one the same quantity for a spheroid having
$Q_0=0.1$, $b_{\rm i}=(-0.5,-0.5,1)$, and the last bottom curve the
velocity for a
spheroid having $Q_0=0.2$, $b_{\rm i}=(-0.5,-0.5,1)$.\\
The figure shows
two trends:\\
a) the collapse velocity is reduced with increasing initial
asymmetry. For example for $a_1/a_3=0.3$ the collapse velocity is
reduced to the Hubble velocity in the plane of collapse (plane of
the pancake for oblate spheroids ) ($H_{\rm c} r_{\rm c}$), while
in the case of more extreme 'flattening' $a_1/a_3=0.125$, the
collapse velocity is reduced by a factor of $\simeq 2.5$ with
respect the previous value. In the case of oblate spheroids, for
this last initial asymmetry, this value
is $\simeq 6$.\\
b) The collapse velocity is reduced with increasing strength of
the external field. In the case of the bottom curve, for
$a_1/a_3=0.125$, the collapse velocity is reduced by a factor of
$\simeq 3$. Similarly to item (a), in the case of oblate
spheroids, for this last initial asymmetry, this value can be
larger than $\simeq 6$.

In other words, the slowing down of the rate of growth of density
contrast produces a lowering of the peculiar velocity in
qualitative and quantitative agreement with BS and Szalay \& Silk
(1983).

The results obtained help to clarify the controversy relative to
the previrialization conjecture. According to
this paper and with WS and BS and in agreement with Hoffman
(1986a) and Bertschinger \& Jain's collapse theorem, it is surely
true that the effect of the shear is to reduce the collapse time
of perturbations. As remarked in the item ``b" of section (3) and in
agreement with WS and BS: {\it The collapse time of perturbations
of fixed initial overdensity, for a fixed background, decreases
with increasing initial asymmetry}. \\
The decrease of the collapse time has such entity to compensate
the effect of the increase of density contrast and collapse
velocity produced by the shear. To be more clear, it is useful to
concentrate on Figs. 3-4. If the collapse of the ellipsoids
occurred at a fixed value of $t$, just like for the spherical
model ($t=0.5 t_{\rm ff}$), the larger increase in density
contrast or velocity produced by the initial asymmetry, in
comparison with the spherical model, should have as a result that
at collapse both the density contrast and velocity should be
enhanced (with respect to the spherical model). But in the
ellipsoidal collapse, the collapse time decreases with increasing
initial anisotropy, and as we have previously seen, in case of initial
axial ratio $8:8:1$ it is given by $t \simeq 0.3 t_{\rm
ff}$. As a consequence, the values of density contrast and
collapse velocity at collapse time are always reduced with respect
the spherical collapse, in agreement with WS, BS, Szalay \& Silk
(1983).
%
%
%
%
%

I must also add that
in the real collapse other effects have an important role, (e.g.,
the effects of small scale substructure). Both Hoffman (1986a) and
Bertschinger \& Jain (1994) results are valid for a fluid element,
which has no substructure by definition, while a small scale
substructure produces a slowing down of the collapse at least
in two ways:\\
1) encounters between infalling clumps and substructure internal
to the perturbation (Antonuccio-Delogu \& Colafrancesco 1994;
Del Popolo \& Gambera 1997; Del Popolo \& Gambera 1999);\\
2) tidal interaction of the main proto-structure with substructure
external to the perturbation (Peebles 1990; Del Popolo
\& Gambera 1998).\\
Moreover, it should be pointed out that, as more small-scale power
is present, the collapse of a perturbation may be slowed down in a
way that could inhibit the effect of shear.\\
%
Similarly to Bertschinger \& Jain (1994), the model presented in
this paper does not take account of the substructure internal to
the system.
%
I, however, recall that the same shortcoming was present in
Peebles (1990): in that paper the substructure was
suppressed, since it adopted an homogeneous Poisson distribution
of particles within the protocluster (Peebles 1990). This limit
has the effect of underestimating the effect of previrialization,
(Peebles 1990). In other words, the slowing down of the collapse
obtained in this paper (similarly to that of Peebles (1990)) are
surely smaller than that we shall find if we had used a system
having internal substructure, as in the above point 1.

 Before concluding, I want to spend a few words on the impact of
the result of the paper on our view of structure formation.\\
The reduction of the rate of growth of overdensity and collapse
velocity has several consequences on structure formation. To
begin with, a first consequence is a change of the mass function,
the two-point correlation function, and the mass that accretes on
density peaks. These last consequences are connected to the effects 
of the shear 
(Audit, Teyssier \& Alimi 1997, 1998).
According to Audit, Teyssier \& Alimi (1997, 1998), the mass function 
depends on two parameters, a density threshold $\delta_{\rm c}$ and a 
shear threshold $\sigma_{\rm c}$. 
According to the previous authors, structures results from a collapse along 
their third principal axis, which is slowed down by the effect of the shear 
(in agreement with our results). 
Therefore on small scales, where the shear is statistically greater, structures 
need on average a higher density contrast to collapse and as a consequence 
the number of objects with $\sigma(M) \leq 1$ \footnote{$\sigma(M)$ is the mass variance}, 
decrease as compared to the collapse 
of the second or first axis, and so the mass function is much below the 
standard Press-Shechter prediction.
Even the two-point
correlation function of galaxies and clusters of galaxies results
strongly modified since the two-point correlation function of the collapsed halos is 
directly connected to the number of objects of a given mass
(see  Peebles 1993; Sheth \& Jain 1997; Del Popolo \& Gambera 1999; Del Popolo et
al. 1999). Another important consequence of the
results described is connected to the value of the density
parameter, $\Omega$. Since the initial anisotropy (internal shear)
and the tidal interaction with external objects (external shear),
slow down the collapse infalling velocity, when using the
spherical infall model we underestimate the value of the density
parameter (Szalay \& Silk 1983; Lee et al. 1986; Taruya \& Soda
2000).
The previous described effects are even larger if the structure acquires angular momentum during evolution (Del Popolo \& Gambera 1999; Del Popolo \& Gambera 2000). In our model, we assumed that  
the principal axes of the external tidal tensor are always
oriented along the principal axes of the mass tensor and this implies 
that the linear angular momentum should be zero (at least from the linear phase to shell-crossing), 
and so while it is legitimate to  
speak about the effects of shear on structure formation, we do not take account of the 
effects of angular momentum acquisition on structure formation, at least before gravitational collapse.
\footnote{However, since the collapse of a protostructure is a violent phenomenon, the conditions of
Kelvin's circulation theorem should be violated (Chernin 1970).
This leads to the acquisition of vorticity by the formation of
shock fronts in the protostructure (pancake), in correspondence of
shell-crossing (Doroshkevich 1970).
Analytical studies by Pichon \& Bernardeau (1999) have also shown that vorticity
generation becomes significant at the scales $3-4 {\rm h^{-1} Mpc}$, and
increases with decreasing scale.} 
%
%
As a result, since tidal forces produces effects similar to that due to shear, 
the quoted limit has the consequence of underestimating the global effect on structure formation
of the interaction of a protostructure 
with the neighboring ones
(see also Del Popolo \& Gambera 1999; Del Popolo \& Gambera 2000). 

Finally, we want to recall, as previously reported, that in the case of an initial spherical configuration our model reduces to Watanabe's. 
In this case, the misalignement condition is veryfied and the sphere can acquire angular momentum.

\section{Conclusions}

I examined the effect of internal and external shear on the
evolution of non-spherical inhomogeneities in a EdS universe. The
study was based upon an approximate analytical solution of the
equation of motions of the axes of the ellipsoid. In the first
part of the paper, I found the analytical solution to the quoted
equations and I compared the result with the numerical solution of
Icke (1973) equations, and in the case of {\it isolated}
ellipsoids with the WS analytical solution. The analytical
approximation is in good agreement with the numerical results both
for {\it isolated} and {\it unisolated} ellipsoids, and it gives a
better approximation to numerical results with respect to WS
analytical solution. The quoted solution was used to study the
effect of shear on the density contrast and peculiar velocity. The results show that:\\
a) The collapse time of perturbations of fixed initial
overdensity, for a fixed background, decreases with increasing
initial asymmetry and strength of the external field. To be more
precise, the evolution of the shortest axis is accelerated with
increasing asymmetry of the structure while the opposite is true
for the longest axis. The effect of a positive tidal force, along
a given axis, is that of slowing down the expansion of the
ellipsoid along this axis and the effect of a negative tidal field
is that of accelerating the expansion of the ellipsoid along that
axis: external and internal shear have a qualitative similar
effect on the evolution of the shortest and longest axes. \\
b) The
difference between the velocities along the longest and shortest
axes is larger for {\it unisolated} ellipsoids than for the {\it
isolated} ones and this difference increases with increasing
strength of the external field. \\
c) While for {\it isolated} ellipsoids or for {\it unisolated}
ones with small values of $|\gamma|$, the axial ratio does not
appreciably change until the perturbation enters a strongly
nonlinear regime, (the self-gravity is dominant), for large values
of $|\gamma|$, the collapse is anisotropic even for not large
values of $\delta$. This means that the collapsing region geometry
is strongly influenced by the external shear: if the external
field is strong enough,
then external shear is dominant (with respect
to internal shear) in shaping the region, in agreement with
Eisenstein \& Loeb (1995), Watanabe(1993).\\
d) Increasing the strength of the external field, the value of the
density contrast increases.\\
e) More asymmetrical initial configurations tend to have, at a
given time, larger values of $\delta$. Then internal and external
shear produce a more rapid evolution of
the density contrast.\\
In order to study the effect of shear on the density contrast at
turnaround and velocity at collapse, I re derived the equations of
the density contrast at turn-around and the velocity at collapse
time of BS model, taking account of both internal and
external shear. The results have shown that:\\
(aa) The values of density contrast and collapse velocity at
collapse time are always reduced with respect the spherical
collapse, in agreement with WS, BS, Szalay \& Silk (1983). \\
(bb) The effects of the slowing down of the collapse obtained in
this paper (similarly to that of Peebles (1990)) are surely
smaller than that we would find if we had used a system having
internal substructure.\\
(cc) The shear has a big impact on our view of structure
formation: \\
(cc1) a first consequence is a change of the mass function, the
two-point correlation function, and the mass that accretes on
density peaks (see also Del Popolo \& Gambera 2000; Audit et al.
1997; Del Popolo \& Gambera 1999; Del Popolo et al. 1999; Peebles
1993).\\
(cc2) Another important consequence of the results described is
connected to the value of the density parameter, $\Omega$. When
using the spherical infall model we underestimate the value of the
density parameter (Szalay \& Silk 1983; Lee et al. 1986; Taruya \&
Soda 2000), since
shear slows down the collapse infalling velocity.\\
Almost all the results obtained with the analytical model were
tested against numerical solutions finding always good agreement
with them.
\begin{acknowledgements}
I would like to thank Prof. E. Recami, E. Nihal Ercan, A. Diaferio,
J. D. Barrow
and Y. Ek\c{s}i for some useful comments. \\
Finally, I would like to thank Bo$\breve{g}azi$\c{c}i University
Research Foundation for the financial support through the project
code 01B304.
\end{acknowledgements}

\end{document}